\documentclass[]{preprint}

\usepackage[style=cell-like-with-doi, citestyle=numeric-comp, sorting=none, backend=biber]{biblatex}
\usepackage{xcolor}

\newcommand{\new}[1]{#1}

\usepackage{booktabs}
\usepackage{tabularx}

\usepackage[raggedright]{titlesec}

\usepackage{cuted}
\usepackage{flushend}

\begin{document}

\title{Answering some questions about structured illumination microscopy}

\author{James D. Manton\textsuperscript{1,*}}

\affil{
    \textsuperscript{1} MRC Laboratory of Molecular Biology, Cambridge, CB2 0QH, UK \\
    \textsuperscript{*} jmanton@mrc-lmb.cam.ac.uk
}

\date{16 November 2021}
\shortauthor{Manton}

\abstract{%
    Structured illumination microscopy (SIM) provides images of fluorescent objects at an enhanced resolution greater than that of conventional epifluorescence wide-field microscopy.
    Initially demonstrated in 1999 to enhance the lateral resolution two-fold, it has since been extended to enhance axial resolution two-fold (2008), applied to live-cell imaging (2009) and combined with myriad other techniques, including interferometric detection (2008), confocal microscopy (2010) and light sheet illumination (2012).
    Despite these impressive developments, SIM remains, perhaps, the most poorly understood `super-resolution' method.
    In this article, we provide answers to the 13 questions regarding SIM proposed by Prakash et al., along with answers to a further three questions.
	After providing a general overview of the technique and its developments, we explain why SIM as normally used is still diffraction-limited.
	We then highlight the necessity for a non-polynomial, and not just non-linear, response to the illuminating light in order to make SIM a true, diffraction-unlimited, super-resolution technique.
    In addition, we present a derivation of a real-space SIM reconstruction approach that can be used to process conventional SIM and image scanning microscopy (ISM) data and extended to process data with quasi-arbitrary illumination patterns.
	Finally, we provide a simple bibliometric analysis of SIM development over the past two decades and provide a short outlook on potential future work.
}

\section*{Introduction}
Structured illumination microscopy (SIM) is a method in fluorescence microscopy which is capable of providing resolutions better than those that can be achieved using uniform illumination in epifluorescence wide-field microscopy.
While first demonstrated, at the turn of the millenium, to enhance only lateral resolution \cite{heintzmann_laterally_1999,gustafsson_surpassing_2000,frohn_true_2000,cragg_lateral_2000}, subsequent developments allow for the enhancement of axial resolution as well \cite{gustafsson_three-dimensional_2008,shao_i5s:_2008}.
SIM has become a popular technique, particularly for live cell imaging \cite{hirvonen_structured_2009,kner_super-resolution_2009,shao_super-resolution_2011}, when better resolution is required as, unlike super-resolution methods such as single molecule localisation microscopy (SMLM) \cite{betzig_breaking_1991,betzig_imaging_2006,hess_ultra-high_2006,rust_sub-diffraction-limit_2006,sharonov_wide-field_2006} and stimulated emission depletion microscopy (STED) \cite{hell_breaking_1994,klar_subdiffraction_1999,klar_fluorescence_2000}, it does not require high laser powers nor special fluorophores.
However, as will be discussed further later, in its normal form it only provides a roughly two-fold enhancement of resolution.

It should be noted that, in 1997, an alternative form of structured illumination microscopy was presented by Neil et al. \cite{neil_method_1997}.
Here the goal was not to improve resolution, but to add optical sectioning to wide-field microscopy.
This form of structured illumination has been commercially successful, with further developments of the technology available from Aurox (Clarity), Andor (Revolution DSD2) and Zeiss (Apotome).
For the purposes of this article, SIM will be taken to correspond only to those methods designed to improve resolution (often referred to as SR-SIM, as opposed to the OS-SIM of Neil et al.).

In this article, we aim to provide a general overview of SIM and answer the 13 questions recently proposed by Prakash et al. \cite{prakash_super-resolution_2021} \new{(sections 1--13)}, along with three more of our own invention \new{(sections 14--16)}.
\new{A similar attempt at answering the 13 questions of Prakash et al. has recently been presented by Heintzmann \cite{heintzmann_answers_2021}.
For some questions, the answers given by Heintzmann are significantly different from those presented here: where this is the case, this will be highlighted in later sections.}
In \prettyref{sec:abbe}, we consider the so-called `diffraction limit' and show that SIM as described here does not break it, even though it provides resolutions beyond those achievable with conventional wide-field imaging.
Following this, in \prettyref{sec:high_na}, we consider whether an impressive demonstration of SIM combined with total internal reflection fluorescence (TIRF) microscopy, which achieves a resolution of \SI{84}{nm}, breaks the diffraction limit.
We then look at the conditions required for SIM to break the diffraction limit in \prettyref{sec:non-linear} and \prettyref{sec:switching}.

In the second part of the manuscript, we show that, despite their prevalence, Fourier methods are not necessarily required to reconstruct images (\prettyref{sec:fourier}) and study the challenges of using SIM in thick samples (\prettyref{sec:deep} and \prettyref{sec:contrast}).
We then argue that image scanning microscopy (ISM), while experimentally a different method, is a form of SIM and show how arbitrary illumination patterns can be used (\prettyref{sec:ism}).
Following this, we study the effects of illumination sparsity in \prettyref{sec:sparsity}.
We note that SIM cannot be used to improve the resolution of transmission microscopy (\prettyref{sec:scattering}) and propose a careful approach to the use of machine learning with super-resolution (\prettyref{sec:machine_learning}).

In the last part of the manuscript, we discuss the practicalities of building a SIM system (\prettyref{sec:cost}) and consider how to combine SIM with SMLM (\prettyref{sec:smlm}).
We then investigate how the axial resolution can be improved beyond a two-fold limit without a non-linear response (\prettyref{sec:axial}), how to combine SIM with light sheet microscopy (\prettyref{sec:light_sheet}) and how quickly SIM images can be acquired (\prettyref{sec:speed}).
Finally, we use a simple bibliometric analysis to track interest in SIM since its inception and conclude with an outlook on future developments (\prettyref{sec:outlook}).

\section{What is super-resolution microscopy and should diffraction-limited linear SIM be classed as `super-resolution'?}\label{sec:abbe}
Super-resolution microscopy refers to a set of techniques that can be used to break `the diffraction barrier', a seemingly unsurmountable goal rewarded with the 2014 Nobel Prize in Chemistry.
The trick in this question is deciding exactly what the diffraction barrier is\dots

There are a number of resolution metrics in use, such as the Rayleigh and Sparrow criteria, that derive from trying to answer the question of how closely two stars can appear while still being distinct in naked-eye astronomical observations \new{\cite{rayleigh_investigations_1879,sparrow_spectroscopic_1916}}.
These criteria are, however, heuristic and have no fundamental physical basis.
Ernst Abbe was the first to propose a physically-motivated limit to the resolution of a light microscope by studying diffraction from a grating \cite{abbe_beitrage_1873}.
He considered using a microscope to image a grating using transmitted light and noted that, in order for the image to have any periodicity, at least two of the diffracted orders from the grating must be captured by the objective.
For head-on illumination, i.e. normal to the grating, he noted that three orders (the \(-1\), 0 and \(+1\) orders) were captured, giving a minimum grating periodicity of:
\begin{equation}
	d_\textrm{head-on} = \frac{\lambda_0}{n \sin \alpha},
\end{equation}
where \(\lambda_0\) is the vacuum wavelength, \(n\) the refractive index of the sample and \(\alpha\) the semiaperture angle (i.e. the maximum angle from the objective's optical axis from which orders can be collected).
For convenience, the product \(n \sin \alpha\) is known as the numerical aperture (NA).

\begin{figure*}
	\centering
	\includegraphics[width=\textwidth]{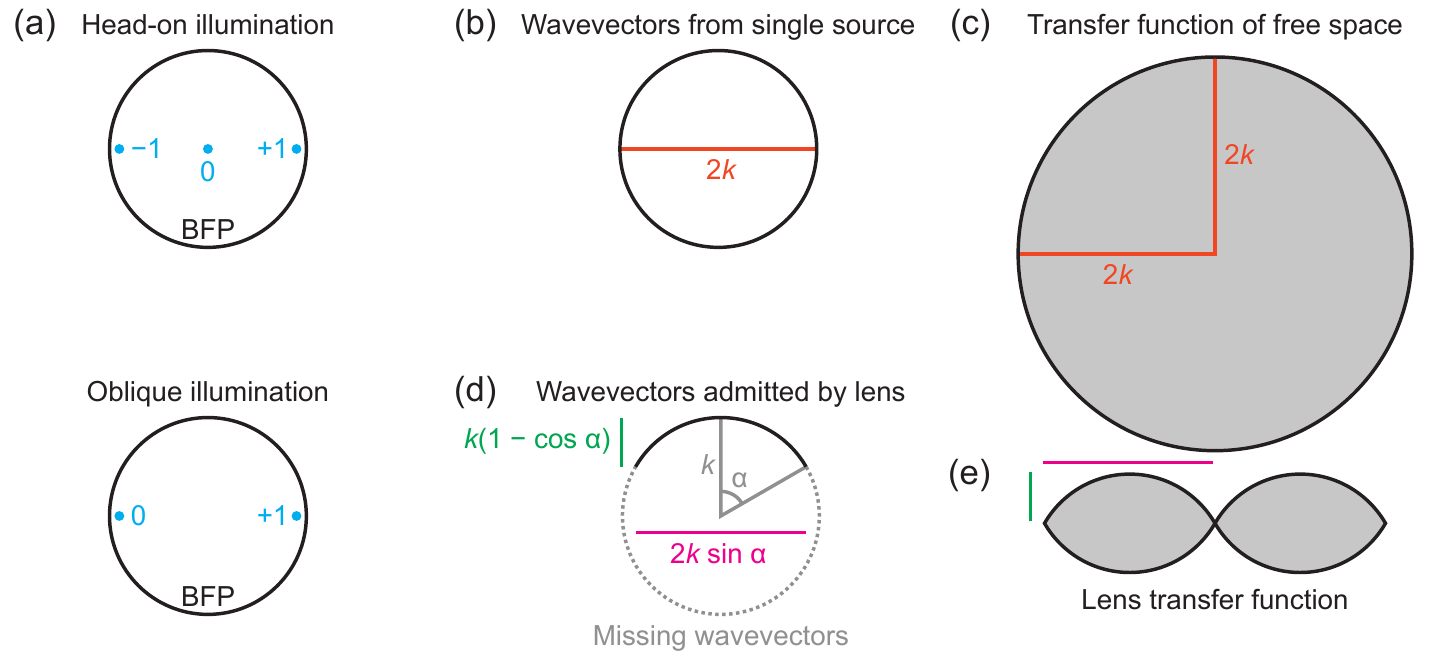}
	\caption{\label{fig:ewald}
		The origin of the diffraction limit.
		(a) Back focal plane (BFP) for head-on and oblique transmission imaging of a grating at the Abbe limits.
		\(\pm 1, 0\) refers to the diffraction orders from the grating.
		(b) Wavevectors from a single source lie on a spherical shell, with a 2D cross-section of this shell shown.
		(c) The transfer function of free space is the autocorrelation of (b), giving a ball of radius \(2k\) (again, shown in cross-section).
		(d) Only a subset of wavevectors are captured by the objective lens, determined by the semiaperture angle, \(\alpha\).
		(e) As with (c), the transfer function is given by the autocorrelation of (d), with the lateral and axial diffraction limits given by the magenta and green geometrical quantities calculated in (d).
	}
\end{figure*}

Referring to \prettyref{fig:ewald}a \new{(top row)}, we see that these three orders are located symmetrically about the centre of the back focal plane.
By tilting the illumination, we can shift the location of these orders laterally.
Given that we only need two orders, not three, for a grating-like structure to be seen, we can imagine that by tilting the illumination such that the zeroth order is right at the edge of the back focal plane, we could increase the order spacing by a factor of two and hence half the grating period (\prettyref{fig:ewald}a\new{, bottom row}).
This leads to the Abbe equation for oblique illumination:
\begin{equation}
	d_\textrm{oblique} = \frac{\lambda}{2n \sin \alpha}.
\end{equation}

\new{What about fluorescence imaging, where light from the sample is emitted in all directions?
Abbe never studied the case of fluorescence imaging, but it turns out that the right hand side of Abbe's equation for oblique illumination is correct for this case as well:}
\begin{equation}
	d^\textrm{lateral}_\textrm{fluorescence} = \frac{\lambda}{2n \sin \alpha}.
\end{equation}
More specifically, this is the lateral resolution, i.e. that within the focal plane.
Perpendicular to the focal plane, the axial resolution is given by
\begin{equation}
	d^\textrm{axial}_\textrm{fluorescence} = \frac{\lambda}{n (1 - \cos \alpha)}.
\end{equation}

\new{Why are these equations correct?}
To explain this, we will follow an argument similar to that of Gustafsson \cite{gustafsson_sevenfold_1995}, which in turn builds on arguments by McCutchen \cite{mccutchen_generalized_1964} and Ewald \cite{ewald_zur_1913}.
First, we must note that we will use the scalar approximation of the electric field, although the argument can be generalised by considering each component of the vectorial electric field separately.
We will also work in the far-field, as near-field components decay exponentially quickly and will not be detected by a normal microscope.
In addition, fluorescence is an incoherent process, with emission from a source bearing no phase relationship to the stimulating light or emission from another source (we ignore the small probability of stimulated emission).
As such, we can limit our consideration to just one source, as the overall response from the sample is the incoherent sum of the responses from all the sources present.

Consider the amplitude of the electric field, \(\new{\mathcal{E}}(\vec{r})\), produced by a source radiating monochromatically.
The source produces a spherical wave which\new{, in the far field,} can be decomposed, via a Fourier transform, into a sum of plane waves with different directions.
All these plane waves will have the same wavelength, \(\lambda\), and so all the wavevectors, \(\vec{k}\), will have the same length.
Hence, as there are no preferred directions and all wavevectors have the same length, the Fourier transform of \(\new{\mathcal{E}}(\vec{r})\), \(\tilde{\new{\mathcal{E}}}(\vec{k})\) \new{(we will use the notation that the Fourier transform of \(X(\vec{r})\) is written as \(\tilde{X}(\vec{k})\) throughout)}, will have the form of a spherical shell, centered on the origin and with radius \(1 / \lambda\) (\prettyref{fig:ewald}b).
\new{This detected far field is very much not the same as the original emitted field, as the evanescent components have been lost.
The consequences of this can most clearly be seen by noting that a true spherical wave would decay smoothly with increasing distance from the origin, while this field has local minima and maxima all over space.}

However, our image is not given by the amplitude of the electric field, but the intensity.
This can be found by multiplying the amplitude by its complex conjugate, giving \(\new{\mathcal{I}} = \new{\mathcal{E}}\new{\mathcal{E}}^*\).
This means that in the Fourier domain, we have
\begin{equation}
	\tilde{\new{\mathcal{I}}}(\vec{k}) = \tilde{\new{\mathcal{E}}}(\vec{k}) \ast \tilde{\new{\mathcal{E}}}^*(-\vec{k}),
\end{equation}
where \(\ast\) denotes convolution.
Hence, \(\tilde{\new{\mathcal{I}}}(\vec{k})\) is the autocorrelation of \(\tilde{\new{\mathcal{E}}}(\vec{k})\) and so will have the form of a ball with twice the radius of the sphere of \(\tilde{\new{\mathcal{E}}}(\vec{k})\), \(2 / \lambda\) (\prettyref{fig:ewald}c).

This is interesting, as it shows that there is a limit to the information carried by the far field, irrespective of the objective lens used: only spatial frequencies up to \(k_c = 2 / \lambda\) are present.
Converting back to real space, this critical spatial frequency corresponds to a \new{spatial} scale of \(\lambda / 2\).
If we consider that our source may be located in a medium of refractive index \(n\), then we can convert \(\lambda\) to a vacuum wavelength, \(\lambda_0 = n \lambda\), and so obtain the far-field limit:
\begin{equation}
	d = \frac{\lambda_0}{2n}.
\end{equation}

Now, let us look at what happens if we use an objective lens to capture the light from our source.
The objective can only capture wavevectors over a certain angular range, meaning that the Fourier domain electric field amplitude distribution has the form of a spherical cap, rather than a full sphere (\prettyref{fig:ewald}d).
As such, the autocorrelation of this has the form of a toroidal solid, shown in cross-section in \prettyref{fig:ewald}e.

We can identify that the maximum lateral spatial frequency is given by \(k_l = 2n \sin \alpha / \lambda_0\), while the maximum axial spatial frequency is given by \(k_a = n(1 - \cos\alpha) / \lambda_0\), where \(\alpha\) is the semiaperture angle.
Converting to real space, this gives:
\begin{equation}
	d_\textrm{lateral} = \frac{\lambda_0}{2n\sin\alpha}
\end{equation}
and
\begin{equation}
	d_\textrm{axial} = \frac{\lambda_0}{n \left(1 - \cos\alpha \right)},
\end{equation}
as was quoted previously.

Considering not just a single source, but a distribution of sources emitting light given by a distribution \(E(\vec{r})\), the toroidal solid formed by autocorrelating the spherical cap defines a transfer function, \(\tilde{H}(\vec{k})\).
Given this, the image spectrum, \(\tilde{D}(\vec{k})\), of the emission spectrum \(\tilde{E}(\vec{k})\) is given by
\begin{equation}\label{eqn:image_formation_fourier}
	\tilde{D}(\vec{k}) = \tilde{E}(\vec{k}) \times \tilde{H}(\vec{k}),
\end{equation}
\new{where \(\times\) denotes point-wise multiplication.
In real space}
\begin{equation}\label{eqn:image_formation_real}
	D(\vec{r}) = E(\vec{r}) \ast H(\vec{r}),
\end{equation}
where \(H(\vec{r})\) is known as the point spread function.

\prettyref{eqn:image_formation_fourier} makes clear that the image spectrum only contains spatial frequencies for which \(\tilde{H}(\vec{k})\) is non-zero.
As the form of \(\tilde{H}(\vec{k})\) is based only on the phenomenon of far-field diffraction, this defines a true diffraction limit \new{based on the spatial frequency bandwidth of \(\tilde{H}(\vec{k})\)}.

Now we are in a position to consider the mechanism of structured illumination and see how it circumvents this limit.
For simplicity, we will restrict our consideration to one dimension and replace \(\vec{r}\) with \(x\).

The fundamental basis of SIM is the realisation that, in our image formation equation, \prettyref{eqn:image_formation_real}, we are not truly interested in the emission distribution function, \(E(x)\), but instead the distribution of sources \(S(x)\).
These distributions are related by \(E(x) = S(x) \times I(x)\), where \(I(x)\) is the illumination distribution.
In the case of uniform illumination, \(I\) merely acts as a scalar multiplier of \(S(x)\).
However, for non-uniform illumination we have an updated image formation equation:
\begin{equation}\label{eqn:sim_real}
	D(x) = E(x) \ast H(x) = \left[S(x) \times I(x)\right] \ast H(x).
\end{equation}
If we consider this image formation equation in the Fourier domain, then we have
\begin{equation}
	\tilde{D}(k) = \tilde{E}(k) \times \tilde{H}(k) = \left[\tilde{S}(k) \ast \tilde{I}(k)\right] \times \tilde{H}(k).
\end{equation}
This suggests that, while the maximum spatial frequency that can be recorded is limited by the support of \(\tilde{H}(k)\), a careful choice of \(\tilde{I}(k)\) may let us shift the sample spectrum, \(\tilde{S}(k)\), around \new{via the convolution} such that other spatial frequencies are transmitted.

SIM, in its simplest variant, achieves this by using illumination with two coherent beams, providing a pattern of the form
\begin{equation}
	I(x) = 1 + \cos(px + \phi),
\end{equation}
or, in the Fourier domain,
\begin{equation}
	\tilde{I}(k) = 2\delta(k) + e^{i\phi}\delta(k + p) + e^{-i\phi}\delta(k - p),
\end{equation}
where \(\delta\) denotes the Dirac delta function.
Hence, for a given phase, \(\phi_m\), we have
\begin{equation}\label{eqn:sim}
	\tilde{D}_m(k) = \left[ 2\tilde{S}(k) + e^{i\phi_m}\tilde{S}(k + p) + e^{-i\phi_m}\tilde{S}(k - p) \right] \times \tilde{H}(k).
\end{equation}
This shows that we have the usual wide-field component, \(\tilde{S}(k)\tilde{H}(k)\), overlaid with two shifted copies of the sample spectrum.
This alone is insufficient to `see' \(\tilde{S}(k \pm p)\), but by recording multiple images with different phases of the illumination pattern, these components can be linearly separated and used to reconstruct an image with improved resolution.
As there are three components to separate, at least three phase shifts are required.

For the purposes of this article, we shall not discuss the details of the image reconstruction routine, nor how SIM can be used to enhance the resolution in all directions (i.e. both laterally and axially).
Instead, the interested reader is directed to the excellent article by Gustafsson et al., which provides detail on both these issues \cite{gustafsson_three-dimensional_2008}.
Further information on the individual steps of reconstruction is presented by Lahrberg et al. \cite{lahrberg_accurate_2018} (pattern estimation), Wicker \cite{wicker_non-iterative_2013} (phase estimation), and Smith et al. \cite{smith_structured_2021} (noise-appropriate recombination).
An overview of the entire reconstruction routine, as well as open-source software, is presented by Lal et al. \cite{lal_structured_2016}.
ImageJ-compatible software for checking the quality of raw and reconstructed data is provided by SIMcheck \cite{ball_simcheck_2015} and for reconstruction by fairSIM \cite{muller_open-source_2016}.

It is clear from \prettyref{eqn:sim} that higher resolutions are achieved with finer illumination periods.
If we illuminate the sample with the same objective lens that we use to capture fluorescence, then the finest illumination we can create is
\begin{equation}
	d_\textrm{illumination} = \frac{\lambda_\textrm{illumination}}{2 n \sin \alpha}.
\end{equation}
Combining this with the \new{emission} diffraction limit, we see that the total resolution is given by
\begin{align}\label{eqn:sim_resolution}
	d_\textrm{SIM} =& \left(\frac{1}{d_\textrm{illumination}} + \frac{1}{d_\textrm{\new{emission}}}\right)^{-1} \nonumber \\
	=& \left( \frac{2 n \sin \alpha}{\lambda_\textrm{illumination}} + \frac{2 n \sin \alpha}{\lambda_\textrm{emission}} \right)^{-1}
\end{align}
and so we have
\begin{equation}
	d_\textrm{SIM} = \frac{\lambda_\textrm{emission}}{\left(\beta + 1\right) 2 n \sin \alpha},
\end{equation}
where \(\beta = \lambda_\textrm{emission} \big/ \lambda_\textrm{illumination}\).
Hence, for the case where \(\lambda_\textrm{emission} = \lambda_\textrm{illumination}\), we see that SIM provides double the resolution of uniform (epifluorescence) illumination.

While this is beyond the resolution of diffraction-limited imaging with uniform illumination, the resolution is still limited, \new{in some sense,} by diffraction.
In order to not be diffraction-limited, we would need to be able to create patterns with arbitrarily fine periods.
Extensions of SIM that \new{provide this} will be discussed in \prettyref{sec:non-linear}.
\new{These methods, theoretically, create images that do not have a spatial band-limit.}
To distinguish SIM, and other techniques, that provide improved resolution but are \new{limited by diffraction in both illumination and emission}, rather than just in emission, the term `extended resolution' has been proposed \cite{gustafsson_extended_1999,chung_extended_2006,stemmer_widefield_2008,li_extended-resolution_2015}.

\new{This allows us to classify methods into four distinct groups based on their theoretical resolving power:
\begin{enumerate}
    \item \emph{Standard resolution} --- as achieved by a normal wide-field epifluorescence microscope; \\
    \item \emph{Extended resolution} --- beyond standard resolution but still limited by the combination of the illumination and emission diffraction limits, as achieved by SIM and ISM (also referred to by some authors as \emph{restricted super-resolution}); \\
    \item \emph{Enhanced resolution} --- beyond extended resolution, including methods with a finite resolution limit, such as photoactivated non-linear SIM; \\
    \item \emph{Super resolution} --- the subset of enhanced resolution methods not limited by diffraction and theoretically capable of infinite resolution, discussed further in \prettyref{sec:non-linear}, such as stimulated emission depletion microscopy and saturated SIM.
\end{enumerate}
Note that some authors, including Heintzmann in his answers to these questions \cite{heintzmann_answers_2021}, would classify all methods capable of imaging beyond the \emph{extended resolution} limit as \emph{super-resolution} methods, but this is not a sufficient condition for `breaking' the diffraction barrier \cite{hell_toward_2003}.
Furthermore, it is worth highlighting the difference between `imaging at the diffraction limit', i.e. with \emph{standard resolution}, and 'diffraction-limited resolution', which may be at any of the levels apart from diffraction-unlimited \emph{super resolution}.
}

\section{Should high-NA TIRF-SIM, which can achieve lateral resolutions down to 84 nm, be considered as diffraction limited?}\label{sec:high_na}
Total internal reflection fluorescence (TIRF) microscopy exploits the nature of total internal reflection to produce a short-range, evanescent illuminating field tethered to a refractive index boundary.
Given two materials with different refractive indices, \(n_1 > n_2\), Snell's law states that light incident on the boundary, from the \(n_1\) side, at angles equal to or greater than \(\arcsin \left(n_2 / n_1\right)\) cannot be transmitted and is instead reflected.
However, Maxwell's equations demand continuity at the boundary and so a rapidly-decaying evanescent field is created on the \(n_2\) side.

For light incident at an angle \(\alpha\), the exponential decay length \new{of the evanescent field} is given by
\begin{equation}
	l = \frac{\lambda}{4\pi \sqrt{n_1^2 \sin^2 \alpha + n_2^2}},
\end{equation}
and is typically of the order of \SI{100}{nm}.
While this ensures that only a thin section of the sample is illuminated, it does not enhance axial resolution as such as the illumination cannot be scanned over the sample.
TIRF-SIM uses two such high-angle beams to (i) reduce background fluorescence, (ii) provide finer patterns than can be created directly in the \(n_2\) medium \cite{fiolka_structured_2008}.

When imaging a watery sample, for which \(n = 1.33\), the maximum possible numerical aperture of a water-immersion objective is \(n\) (i.e. \(\sin \alpha = 1\)), while oil immersion objectives with numerical apertures of 1.4, 1.49, 1.5 and even 1.7 are available.
These objectives can form a pattern right at the coverslip boundary at the limit of their numerical apertures and have the near-field be patterned with the same spatial period, even though there is no way of creating such a fine pattern in water using far-field illumination.

Li et al. used such a 1.7-NA objective with between 30--\SI{100}{W . cm^{-2}} of \SI{488}{nm} illumination at 1.58 NA and \textasciitilde\SI{510}{nm} detection and claim a resolution of \SI{84}{nm} \new{\cite{li_extended-resolution_2015}}.
Using a modified version of \prettyref{eqn:sim_resolution}, where we allow the numerical apertures of illumination and detection to be different, we can back-calculate the effective detection NA for their claim to be 1.38, consistent with the high end of the range of reported cytosolic refractive indices \cite{liu_cell_2016}.
The corresponding non-SIM TIRF resolution for this case would then be \(2.2\times\) worse, at \SI{185}{nm}, while the maximum possible far-field SIM resolution would be \SI{90}{nm}.
Nevertheless, this \SI{84}{nm} is still diffraction-limited, as the period of the pattern created at the boundary is still limited by diffraction on the \(n_1\) side.
It is interesting to note that the authors rejected the possibility of collecting super-critical fluorescence (i.e. near-field emission that is converted to propagating light by interacting with the boundary) to further enhance their resolution calculation (i.e. the detection NA is limited to the refractive index of the sample), despite other measurements with the same objective showing that significant super-critical light can be present \cite{dasgupta_direct_2021}.

\new{It is particularly instructive to compare this answer (`yes') with the companion given by Heintzmann: "a clear `no'" \cite{heintzmann_answers_2021}.
For Heintzmann, taking \prettyref{eqn:sim_resolution}, all forms of SIM break the emission Abbe limit and hence break the diffraction limit.
This equation can be used to define a new limit, that of `restricted super resolution', introduced by Sheppard \cite{sheppard_fundamentals_2007}.
For TIRF-SIM, the structured illumination pattern has a finer frequency than that permitted by the refractive index of the sample, and so Heintzmann says that TIRF-SIM breaks this `restricted super resolution' criterion.
Hence, the suggestion is that TIRF-SIM is not diffraction-limited even under the considerations of this more strict approach.

However, the fact remains that the fineness of the illumination pattern is limited by diffraction, just in a medium of refractive index \(n_1\) rather than \(n_2\).
Even in the situation where the sample was mounted on a diamond coverslip (ignoring the deleterious effects of the strong dispersion in such a material) and we had an objective lens with numerical aperture matching that of the refractive index of the material, 2.4, TIRF-SIM of the same sample as Li et al. used could not produce a resolution better than
\begin{equation}
    d = \left( \frac{2 n_1 \sin \alpha}{\lambda_\textrm{illumination}} + \frac{2 n_2 \sin \alpha}{\lambda_\textrm{emission}} \right)^{-1} = \left( \frac{2 \times 2.4}{\SI{488}{nm}} + \frac{2 \times 1.38}{\SI{510}{nm}} \right)^{-1} = \SI{65.6}{nm}.
\end{equation}
Hence, this author would class this as \emph{extended resolution}, not \emph{super resolution} or even \emph{enhanced resolution}.
}

\section{Can non-linear SIM become broadly applicable and live-cell compatible?}\label{sec:non-linear}
Non-linear SIM breaks the assumption that the fluorescent response from the sample is linearly related to the intensity of the illuminating light \new{(\(E(x) \neq S(x) \times I(x)\))}.
The non-linearity adds extra harmonics (i.e. spatial frequency components above that of the linear case) to the effective illumination pattern, further extending the resolution.
\new{A patent application (DE19908883A1) describing this idea was submitted by Heintzmann and Cremer in March 1999, barely two months after the publication of the first SIM paper.}
The simplest case to consider is a normal SIM system in which a further sinusoidal pattern, \(R(x) = 1 + \cos(px + \phi)\), is used to control the fluorescence response\new{, e.g. by photoswitching a fraction of the molecules between dark and light states}.
Now, the camera data acquired have the form:
\begin{equation}
	D(x) = \left[ I(x) \times R(x) \times S(x) \right] \ast H(x),
\end{equation}
and in the Fourier domain:
\begin{equation}
	\tilde{D}(k) = \left[ \tilde{I}(k) \ast \tilde{R}(k) \ast \tilde{S}(k) \right] \times H(x).
\end{equation}

\begin{figure*}
	\centering
	\includegraphics[width=\textwidth]{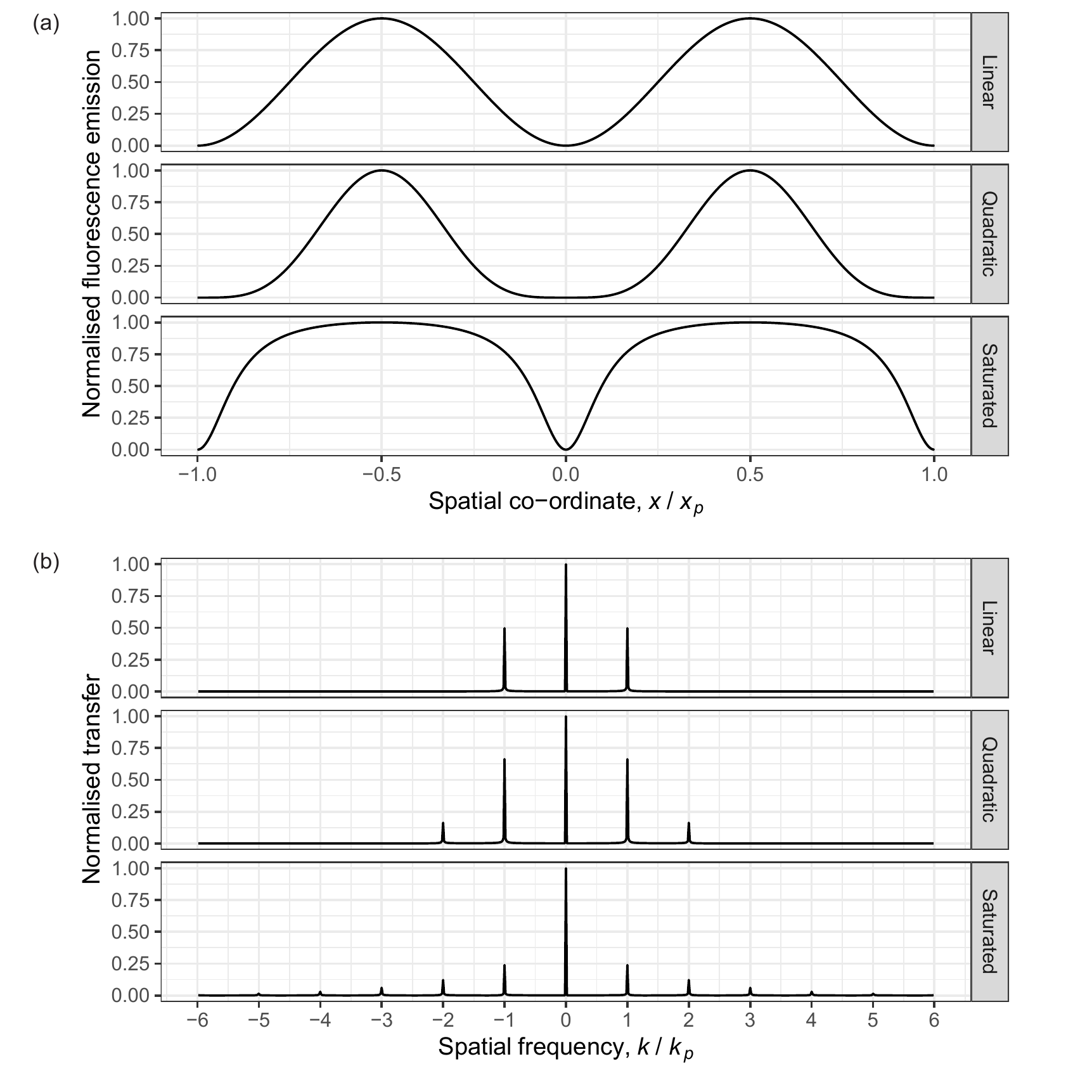}
	\caption{\label{fig:non-linear}
		Real space (a) and Fourier domain (b) comparisons of linear SIM (top row), non-linear SIM with a quadratic non-linearity (middle row), and non-polynomial SIM with fluorescence saturation (bottom row).
		The pattern period and spatial frequency are denoted by \(x_p\) and \(k_p\), respectively.
		Note that, in addition to containing an extra harmonic at \(k = \pm 2 k_p\), the quadratic case has a stronger component at \(k = \pm k_p\) than in the linear case.
		Furthermore, while harmonics beyond the plot range are present in the saturated fluorescence case, their relative strengths compared to the DC peak at \(k = 0\) are so weak that the harmonic at \(k = \pm k_p\) is already barely visible.
		The saturated fluorescence case was calculated assuming a \new{mean excitation irradiance of 1 photon per cross-section per lifetime} and continuous wave excitation.
	}
\end{figure*}

Assuming that \(I(x)\) and \(R(x)\) have the same period, \(p\), we now have a total of five delta functions, at \(k \in \left\{-2p, -p, 0, p, 2p\right\}\) (see \prettyref{fig:non-linear}b middle row).
By phase-shifting both \(I(x)\) and \(R(x)\) simultaneously, we can unmix these five components with only five images, producing a reconstruction with sample information at three times the conventional limit.

Despite the further resolution increase, even this approach is diffraction-limited, as \(I(x)\) and \(R(x)\) are both individually diffraction-limited.
\new{Hence, the spatial frequencies available through this method are limited to a certain band.}
In effect, all we have done is taken the normal SIM equation (\prettyref{eqn:sim_real}) and replaced \(I\) with \(I_\textrm{eff} = I^2\).
We could imagine \new{hypothetically} extending the scheme even further, with yet another sinusoidal pattern controlling the response of the sample to the light which controls the fluorescence response, but this would correspond to \(I_\textrm{eff} = I^3\) and so, while providing higher resolution (to four times the usual limit), would still be diffraction-limited.
\new{Hence, while these methods would provide \emph{enhanced resolution} beyond \emph{extended resolution}, they do not belong to the class of diffraction-unlimited \emph{super-resolution} methods.}

In order to achieve true \new{super-resolution} imaging with SIM, we need not just a non-linear response, but a non-polynomial one.
\new{This is because a polynomial response can be considered as a (scaled, repeated) convolution of the illumination spectrum with itself and so, as we start with a finite number of harmonics, we must end with a larger, but still finite, number.
In contrast, a non-polynomial response} would provide an infinite number of harmonics and hence cannot be said to be limited by diffraction (see \prettyref{fig:non-linear}a,b bottom rows).

\new{To recap, a uniform illumination pattern provides \emph{standard resolution} at the diffraction limit.
Patterned illumination with a linear fluorescence response to illumination light provides \emph{extended resolution} up to twice the diffraction limit (assuming excitation and emission light have the same wavelength).
Replacing the linear response with a non-linear, polynomial response provides further \emph{enhanced resolution}, albeit one that is still bandlimited.
Finally, implementing a non-polynomial response allows for diffraction-unlimited \emph{super-resolution} imaging.}

The first non-polynomial SIM method to be proposed and demonstrated involves saturating the fluorescence, such that a linear increase in illumination intensity results in a sub-linear increase in fluorescence \cite{heintzmann_saturated_2002} (compare the top and bottom rows of \prettyref{fig:non-linear}a).
Saturation occurs when a significant fraction of the illuminated molecules are in the excited state when further excitation photons are incident, as there are insufficient ground-state molecules that can be excited by the photon flux.
So far, the only demonstration of saturated SIM was provided by Gustafsson in 2005 \cite{gustafsson_nonlinear_2005}, which showed the ability to resolve a close-packed monolayer of \SI{50}{nm}-diameter fluorescent beads.
Here, pulsed illumination was used to provide an \(I_\textrm{eff} \propto 1 - \exp\left[-(I \sigma + k_f) t_p\right] \), where \(\sigma\) is the fluorescence excitation cross-section of the fluorophore used, \(k_f = 1 / \tau\) (where \(\tau\) is the fluorescence lifetime) and \(t_p\) is the pulse duration.

While, theoretically, an infinite number of harmonics were present, the signal-to-noise ratio limited the number of `usable' harmonics to three, i.e. a five-fold resolution enhancement, although five harmonics were present in calibration data.
\new{This is a good illustration of the point that, while \emph{super-resolution} methods may theoretically be capable of providing superior resolution to \emph{enhanced-resolution} methods, in practice the distinction is often irrelevant.}

Given that the average illumination irradiance was only \SI{35}{mW . cm^{-2}}, one may wonder why this method has never been demonstrated on biological samples.
This is because considering only the average irradiance is misleading, as the illumination was pulsed.
Averaging instead over the pulse duration gives a peak irradiance of \SI{10}{MW . cm^{-2}}, comparable to the levels used for multiphoton microscopy.
Photodamage mechanisms are known to be non-linear with respect to light intensity, such that a doubling in intensity produces more than double the damage.
Nevertheless, it is not clear that this irradiance is necessarily too damaging for imaging biological samples, especially as the saturation irradiance of EGFP has been measured to be similar, at \SI{11}{kW . cm^{-2}} \cite{kubitscheck_imaging_2000}, and multiphoton microscopy has successfully been used to image many varied biological samples.
\new{While multiphoton microscopy uses redder wavelengths of illumination, which is suggested to contribute less to photodamage, standard use of confocal fluorescence microscopy also often operates at illumination irradiances close to the saturation irradiances of similar fluorophores.}

Another non-polynomial SIM approach, published in 2012 by Rego et al., involved saturated depletion of fluorescence, in which the photoswitchable nature of the fluorescent protein Dronpa was used to confine fluorescence to thin bands \cite{rego_nonlinear_2012}.
\new{This followed a 2008 demonstration of the same mechanism by Hirvonen et al., although here the authors were unable to successfully reconstruct a high-resolution image \cite{hirvonen_structured_2008}.}
\new{The idea of saturating the depletion of fluorescence is one example of the RESOLFT concept \cite{hell_toward_2003}, which was demonstrated in a parallelised, wide-field manner in 2007 \cite{schwentker_wide-field_2007}.}
\new{In the work of Rego et al., images with} resolutions better than \SI{50}{nm} were produced despite the required irradiances being of the order of only \SI{10}{W . cm^{-2}}.
Unfortunately, only fixed cells were imaged as one reconstructed frame required \SI{945}{s} of imaging time.

In order to develop a live-cell compatible method, Li et al. abandoned the use of non-polynomial SIM and instead focussed on developing a technique that would work without the need for saturation \cite{li_extended-resolution_2015}.
To this end, they used the photoswitchable fluorescent protein Skylan-NS for patterned activation, rather than depletion.
This enabled the timelapse imaging of live cells at resolutions better than the \SI{84}{nm} of their high-NA linear SIM approach (see \prettyref{sec:high_na})\new{, producing a movie of 30 frames at \SI{65}{nm} resolution}.
Furthermore, by increasing the activation irradiance to \SI{490}{W . cm^{-2}}, they began to saturate the transition, creating a non-polynomial method with an extra harmonic above the noise floor.
\new{This increased the resolution to \SI{45}{nm}, but only provided 12 frames at useful signal-to-noise ratios.}

\new{Hence, both non-linear and non-polynomial SIM have been demonstrated to be live-cell compatible.
However, while linear SIM can image live samples for hundreds of time points, non-linear approaches are currently limited to a few tens of time points.
In addition, while non-polynomial methods theoretically provide unlimited resolution, signal-to-noise concerns show that they are, in practice, less well-suited to extending resolution than carefully chosen non-linear methods such as the patterned activiation of Li et al..
As an example of this, compare the `quadratic' and `saturated' plots in \prettyref{fig:non-linear}b.
While the saturated case has more harmonics in total, the strengths of those at \(k_p\) and \(2k_p\) are much stronger in the quadratic case.
Given a finite photon budget, more photons must be spent to raise these harmonics above the spectrally-flat noise floor using saturation.
This problem is exacerbated at higher illumination powers as, while the strength of higher harmonics increase, the strength of the first harmonic decreases (see \prettyref{svid:ssim}).

Numerical simulations suggest that saturation using continuous wave illumination, for which an illumination of \(I(x)\) is mapped to an effective excitation of \(I / (I + 1 / \tau)\), produces harmonics that decay geometrically beyond the first harmonic, with the strength of the first harmonic decaying with increased saturation strength.
For pulsed illumination, a similar pattern of harmonics is seen, with a short pulse length mimicking the effect of a higher intensity.
For polynomial illumination response with \(N\) harmonics (i.e. \(2N + 1\) peaks in the spectrum) and perfect modulation contrast, the strength of the \(n\)\textsuperscript{th} harmonic is given by
\begin{equation}
    s = \frac{(N!)^2}{(N + n)! (N - n)!},
\end{equation}
where \(s = 1\) would correspond to a harmonic as strong as the DC peak (\(n = 0\)).
These strengths can be read off from every other row in Pascal's triangle (dividing by the central element), with uniform illumination corresponding to the first row.}

\section{Do you need `switching' of states for non-linear super-resolution imaging?}\label{sec:switching}
So far, all of our discussions of non-linear SIM and super-resolution imaging have used continuous models.
Despite this, we know that light is quantised into photons and fluorescent samples are quantised into fluorophores.
As such, all the approaches discussed in \prettyref{sec:non-linear} feature `switching' of states, but this is also true of normal fluorescence imaging.
\new{Hence, switching is not a sufficient condition for non-linear super-resolution microscopy.}
While switching of states can provide useful non-linearities, it alone is not sufficient to encompass all possible non-linearities (e.g., Rabi oscillations \cite{heintzmann_subdiffraction_2009}), merely those that have been successfully demonstrated to work with SIM so far.
\new{Hence, switching is theoretically not a necessary condition for non-linear super resolution, but has been utilised in all methods implemented so far.
For more details on alternative super-resolution mechanisms, see Heintzmann's answer in reference \cite{heintzmann_answers_2021}.}

\section{Do high-quality SIM images require reconstruction in Fourier space?}\label{sec:fourier}






In general, SIM reconstructions can be split into two separate steps: (i) estimating the parameters of the pattern of illumination used (period, orientation and phase) and (ii) using these parameters to separate and recombine information contained within the raw images.
\new{In many implementations}, both of these steps have been conducted in the Fourier domain, although there is no inherent reason why they must be.

Pattern estimation can potentially be skipped if the system is well-calibrated and the sample does not distort the illumination pattern.
If the pattern parameters are unknown, they are typically acquired via cross-correlating images with different phases.
Assuming the pattern parameters are well-known, either via calibration or through a pattern estimation routine that processes the raw data, a typical reconstruction would separate the different information components, shift them to their proper locations in Fourier space, and recombine them to form an image.

\new{Separating the information components is typically thought of as solving the Fourier space matrix equation
\begin{equation}
    \begin{bmatrix}
        \tilde{D}_1(k) \\
        \tilde{D}_2(k) \\
        \tilde{D}_3(k) \\
    \end{bmatrix}
    =
    \underbrace{\begin{bmatrix}
        1 & 1 & 1 \\
        1 & e^{2\pi i/3} & e^{-2\pi i/3} \\
        1 & e^{4\pi i/3} & e^{-4\pi i/3} \\
    \end{bmatrix}}_{M}
    \begin{bmatrix}
        2 \tilde{H}(k) \tilde{S}(k) \\
        \tilde{H}(k) \tilde{S}(k + p) \\
        \tilde{H}(k) \tilde{S}(k - p) \\
    \end{bmatrix},
\end{equation}
where \(M\), the mixing matrix, has been calculated for the case of equal \(2\pi/3\) phase steps with no global phase offset, for simplicity.
However, this mixing matrix acts pixel-wise and so there is no reason to not consider the real space equivalent instead:
\begin{equation}
    \begin{bmatrix}
        D_1(x) \\
        D_2(x) \\
        D_3(x) \\
    \end{bmatrix}
    =
    \underbrace{\begin{bmatrix}
        1 & 1 & 1 \\
        1 & e^{2\pi i/3} & e^{-2\pi i/3} \\
        1 & e^{4\pi i/3} & e^{-4\pi i/3} \\
    \end{bmatrix}}_{M}
    \begin{bmatrix}
        2 H(x) \ast S(x) \\
        H(x) \ast S(x) e^{2\pi i px} \\
        H(x) \ast S(x) e^{-2\pi i px} \\
    \end{bmatrix}.
\end{equation}
This particular mixing matrix can be seen as a discrete Fourier transform along the phase direction, but this is not generally true and is not what is meant by a reconstruction in Fourier space (as this operates along the spatial directions).
If we allow our image processing routine to use complex numbers then we can separate each component by calculating the matrix inverse of \(M\) and applying it to the raw data.

Shifting each component to the correct location in Fourier space can be achieved in real space, via the Fourier shift theorem, by multiplying with a phase ramp \new{(see \prettyref{svid:shifting})}.
It should, however, be noted that in many reconstructions a Fourier transform and inverse transform of a padded spectrum are used to artificially reduce the pixel size before shifting, as the raw data are not Nyquist-sampled for the higher resolution reconstruction.
Staying in real space either requires oversampled raw data or an alternative upscaling technique.

Given shifted components, all that remains is to combine them to form a SIM image.
This can be achieved with a direct sum, but most reconstructions choose to weight different parts of the separated components differently depending on their exact location in Fourier space, in an attempt to counteract the effects of the OTF.

As an alternative to this view of separating and recombining components,} we can think of the process of structured illumination imaging as akin to amplitude modulation (AM), as used in early radios.
Here, the carrier signal is our illumination pattern, the sample structure is our message signal and the recorded data is the product of these two, as is the AM signal.
To demodulate an AM signal and recover the message signal, a radio can use a `product detector', which heterodynes (i.e. multiplies) the AM signal with a signal from a local oscillator designed to match the carrier signal exactly.
This suggests that a similar approach might be applicable to SIM data.

Consider a two-beam SIM system where, for simplicity, we will only consider one pattern orientation and perfect modulation contrast.
In the usual way, the data acquired, \(D_m(x)\), where \(m\) is used as a phase index, are equal to the product of the illumination and sample, convolved with the PSF, \(H(x)\):
\begin{align}
	D_m(x) &= \left[ I(x) \times S(x) \right] \ast H(x) \\
	&= \left[ \left(1 + \cos(px + \phi_m)\right) \times S(x) \right] \ast H(x).
\end{align}
As usual, in the Fourier domain:
\begin{equation}
	\tilde{D}_m(k) = \bigg[ 2 \tilde{S}(k) + e^{i\phi_m}\tilde{S}(k + p)  + e^{-i\phi_m}\tilde{S}(k - p) \bigg] \tilde{H}(k).
\end{equation}

Let us now consider multiplying our AM signals, \(D_m(x)\), with a `local oscillator', i.e. another sinusoidal pattern with the same phase and period:
\begin{equation}
	M_m(x) = \alpha + \beta\cos(px + \phi_m).
\end{equation}
In the Fourier domain, our local oscillator takes the form:
\begin{equation}
	\tilde{M}_m(k) = \alpha \delta(k) + \beta e^{i\phi_m}\delta(k + p)  + \beta e^{-i\phi_m}\delta(k - p).
\end{equation}
Now, let us heterodyne our signals:
\begin{equation}
	P_m(x) = D_m(x) \times M_m(x),
\end{equation}
such that in the Fourier domain we have:
\begin{equation}
	\tilde{P}_m(k) = \tilde{D}_m(k) \ast \tilde{M}_m(k).
\end{equation}
Calculating this gives:
\begin{align}
	\tilde{P}_m(k) =& 2\alpha\tilde{S}(k)\tilde{H}(k) \nonumber \\
	& + \alpha e^{i\phi_m}\tilde{S}(k + p)\tilde{H}(k) + \alpha e^{-i\phi_m}\tilde{S}(k - p)\tilde{H}(k) \nonumber \\
	& + 2\beta e^{i\phi_m}\tilde{S}(k + p)\tilde{H}(k + p) \nonumber \\
	& + 2\beta e^{2i\phi_m}\tilde{S}(k + 2p)\tilde{H}(k + p) + \beta\tilde{S}(k)\tilde{H}(k + p) \nonumber \\
	& + 2\beta e^{-i\phi_m}\tilde{S}(k - p)\tilde{H}(k - p) \nonumber \\
	& + \beta\tilde{S}(k)\tilde{H}(k - p) + 2\beta e^{-2i\phi_m}\tilde{S}(k - 2p)\tilde{H}(k - p).
\end{align}

If we impose the conditions that
\begin{equation}
	\sum_m e^{i\phi_m} = 0 \quad \& \quad \sum_m e^{2i\phi_m} = 0
\end{equation}
then most of these terms cancel when summed, giving:
\begin{align}
	\sum_m \tilde{P}_m(k) =& 2m\alpha\tilde{S}(k)\tilde{H}(k) \nonumber \\
	& + m\beta\tilde{S}(k)\tilde{H}(k + p) + m\beta\tilde{S}(k)\tilde{H}(k - p).
\end{align}
Hence, the reconstructed image, obtained by multiplying each raw image with our local oscillator and summing all the images, has the same form as a traditional SIM reconstruction if we set \(\alpha = \beta = 1\).
This form also makes clear why a SIM acquisition with only two phases does not reconstruct properly: in this case \(\sum_m e^{i\phi_m} = 0\) as desired, but \(\sum_m e^{i\phi_m} = 2\) and so we are left with terms of the form \(\tilde{S}(k \pm 2p)\tilde{H}(k \pm p)\) that we do not want.

\begin{figure*}
	\centering
	\includegraphics[width=\textwidth]{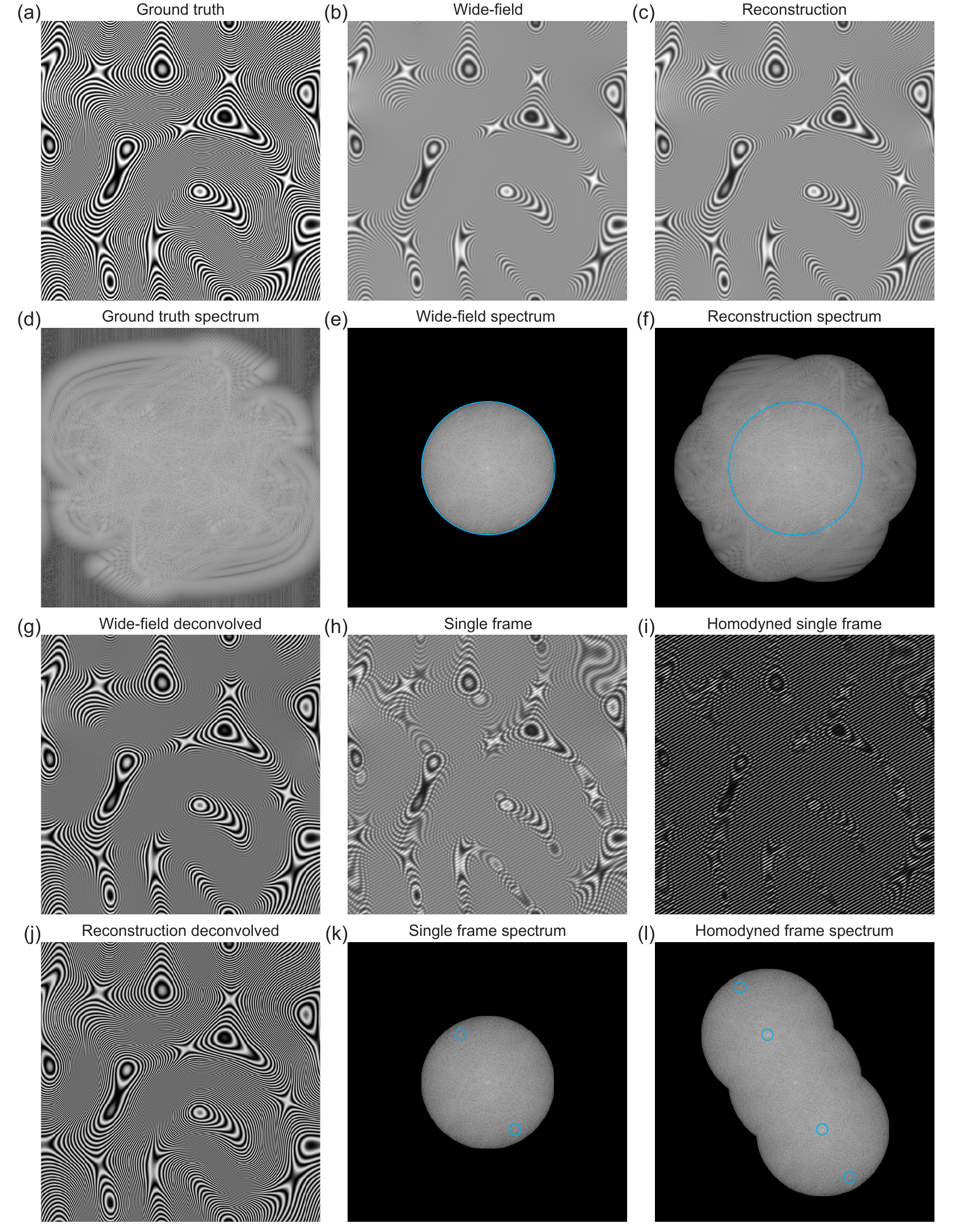}
	\caption{\label{fig:real-space}
		An overview of a reconstruction using the real-space method described in the main text.
		(a) Ground truth sample structure.
		(b) Wide-field image obtained with uniform illumination.
		(c) Complete SIM reconstruction.
		(d,e,f) Logarithmically-scaled Fourier spectra of (a,b,c).
		(g) \new{Linearly} deconvolved wide-field image.
		(h) Single frame of SIM acquisition.
		(i) Frame from (h) multiplied by local oscillator.
		(j) \new{Linearly} deconvolved SIM reconstruction.
		(k) Logarithmically-scaled Fourier spectrum of (h).
		Peaks circled in cyan correspond to the DC peaks of \(\tilde{S}(k \pm p)\) in \(\tilde{D}_m(k)\).
		(l) Logarithmically-scaled Fourier spectrum of (i).
		Peaks circled in cyan correspond to the undesired DC peaks of \(\tilde{S}(k \pm p)\) and \(\tilde{S}(k \pm 2p)\) in \(\tilde{P}_m(k)\).
		Note that these peaks are not present in (f) as they are annihilated when all phases \(m\) are summed over \new{(the illumination peaks are deliberately chosen to be well inside the cyan line, for clarity)}.
	}
\end{figure*}

\new{\prettyref{fig:real-space} shows an example of this reconstruction process for simulated data.
Here, the ground truth sample (\prettyref{fig:real-space}a) was engineered to be periodic, to avoid `cross' artefacts in the spectra (\prettyref{fig:real-space}d,e,f).
In addition, its form makes the Moir\'e effect of the fringe-like illumination particularly visible (\prettyref{fig:real-space}h).}

Considering a three-beam system, for which
\begin{align}
	\tilde{D}_m(k) = & \bigg[ 3 \tilde{S}(k) + 2e^{i \phi_m}\tilde{S}(k + p)  + 2e^{-i\phi_m}\tilde{S}(k - p) \nonumber \\
	& + e^{2i\phi_m}\tilde{S}(k + 2p)  + e^{-2i \phi_m}\tilde{S}(k - 2p) \bigg] \tilde{H}(k)
\end{align}
and
\begin{align}
	\tilde{M}_m(k) = & \alpha \delta(k) + \beta e^{i\phi_m}\delta(k + p)  + \beta e^{-i\phi_m}\delta(k - p) \nonumber \\ 
	& + \gamma e^{2i\phi_m}\delta(k + 2p)  + \gamma e^{-2i \phi_m}\delta(k - 2p),
\end{align}
this approach provides
\begin{align}
	\frac{1}{m}\sum_m\tilde{P}_m(k) = & 3\alpha\tilde{S}(k)\tilde{H}(k) \nonumber \\
	& + 2\beta\tilde{S}(k)\tilde{H}(k + p) + 2\beta\tilde{S}(k)\tilde{H}(k - p) \nonumber \\
	& + \gamma\tilde{S}(k)\tilde{H}(k + 2p) + \gamma\tilde{S}(k)\tilde{H}(k - 2p)
\end{align}
and so we see that we should set \(\alpha = \beta = \gamma = 1\).
Indeed, for \(n\) beams, the correct \(\tilde{M}_m(k)\) is one which is spectrally flat, i.e. all delta functions have equal amplitude.
This reconstruction approach will be revisited in \prettyref{sec:ism} to consider the case of non-sinusoidal illumination.
\new{Similar real-space reconstruction approaches were also considered by Cragg \& So early-on in the development of SIM \cite{cragg_lateral_2000,so_resolution_2001}.}

\new{Intriguingly, the idea of using a patterned illumination and a local oscillator to enhance the resolution of an imaging system predates the idea of structured illumination microscopy by more than 30 years.
In a series of papers, beginning in 1966, Lukosz investigated the possibility of enhancing the resolution of an imaging system by inserting opaque masks into object and image space \cite{lukosz_optical_1966,lukosz_optical_1967,bachl_experiments_1967}.
While the details of the method are quite different to the modern idea of structured illumination microscopy, the underlying idea and the realisation that an increase in spatial bandwidth must be accompanied by a decrease in temporal bandwidth were already present.
}

\section{Can SIM be used for deep tissue imaging?}\label{sec:deep}
Using SIM for deep tissue imaging is complicated by two issues: (i) sample-induced aberrations and (ii) insufficient stripe contrast caused by out-of-focus and scattered light.
We will restrict the discussion here to that of sample-induced aberrations and discuss (ii) in \prettyref{sec:contrast}.
In larger samples, it is common for the local refractive index to not be uniform over the entire sample.
This means that light that arrives at or departs from a given location in the sample experiences a path-dependent phase shift with respect to light travelling in a uniform refractive index.
These unanticipated phase shifts mean that the light is not properly focussed by the microscope system.
This is true for both illumination light and fluorescence emission.

Fortunately, it has been shown by Arigovindan et al. that aberrations predominantly affect the fluorescence emission light, as the effect of phase shifts on the illumination pattern can be taken into account in the reconstruction routine \cite{arigovindan_effect_2012}.
However, this assumes that the variation across the field-of-illumination is small (i.e. the `isoplanatic patch' is large), otherwise the image must be split into blocks over which the pattern parameters are fixed.
Correcting for aberrations in the fluorescence emission can be achieved using an `adaptive optic', typically a deformable mirror, which applies a complementary phase shift to each part of the wavefront to cancel the aberrations.
Usefully, this \new{can not} only be used to compensate for sample-induced aberrations, but any residual aberrations from the microscope itself.
However, as the same correction is applied to every point in the field-of-view there is once again a requirement that the variation across the field is small.
This is a limitation of all wide-field adaptive optics systems and is in constrast to point-scanning systems, for which the aberration correction can be updated for every scan position, if necessary \cite{booth_adaptive_2014}.

Such adaptive optics SIM systems have been constructed by a number of groups, with most differences being in the way an appropriate wavefront correction is obtained.
Using a multiphoton guide star allows for direct wavefront sensing, in which the wavefront is measured using a Shack--Hartmann sensor, but \new{typically} requires an expensive femtosecond laser which may cause excess sample damage \cite{turcotte_dynamic_2019,lin_structured_2021}.
In contrast, sensorless schemes assume that the wavefront can be decomposed into a sum of orthogonal modes, and simply see which combination of modes maximise an image quality metric.
Recently, \v{Z}urauskas et al. exploited the nature of the structured illumination itself to futher improve the performance of such a sensorless scheme \cite{zurauskas_isosense_2019}.
These approaches have the benefit of not requiring a guide star, but mean that a large number of images must be acquired to obtain a good correction (at best \(2N + 1\) for \(N\) modes) \cite{booth_adaptive_2014}.

\section{How can the fundamental limitation of SIM, i.e. generating sufficient stripe contrast in densely labelled and/or extended biological structures due to out-of-focus light, be addressed?}\label{sec:contrast}

Even for situations in which the adaptive optics schemes discussed in \prettyref{sec:deep} are not required, obtaining high-quality SIM reconstructions can be challenging in thicker samples due to the out-of-focus light contaminating the image and hence reducing stripe contrast.
For single-cell imaging, the out-of-focus light produced by 2-beam or 3-beam SIM is normally insufficient to preclude high-quality reconstructions.
However, for tissue imaging, background fluorescence from scattered light and out-of-focus regions of the sample can quickly degrade imaging performance.

In general, there are two approaches to solve this problem: (i) avoid producing out-of-focus light in the first place; (ii) somehow remove the out-of-focus light before it hits the detector.
In conventional fluorescence microscopy, (i) is the approach taken by multiphoton and light sheet microscopy, while (ii) is that taken by confocal microscopy.
All three approaches can usefully be combined with SIM.

In 2012, Andresen \& Pollok et al. created a sinusoidal multiphoton SIM pattern by scanning an array of 32 beamlets perpendicular to the beamlet line, with phase-stepping and reconstruction in the usual manner \cite{andresen_high-resolution_2012}.
However, as each beamlet is incoherent with the others, the fine fringe patterns necessary for a high resolution enhancement would have too low a modulation contrast in this scheme, precluding a full doubling of resolution.
Later, various authors combined multiphoton excitation with some form of image scanning microscopy (discussed further in \prettyref{sec:ism}), which uses a focussed spot for excitation and a different reconstruction scheme \cite{winter_two-photon_2014}.
This has also been combined with adaptive optics to improve the imaging performance at depth further still \cite{zheng_adaptive_2017}.

Initially, light sheet microscopy was first combined with incoherent OS-SIM to enhance optical sectioning, rather than resolution \cite{keller_fast_2010}.
Soon after, Planchon et al. used an incoherent superpositions of Bessel beams to provide enhanced resolution as well \cite{planchon_rapid_2011}.
However, this still used the reconstruction approach of OS-SIM, resulting in a spatially-varying point spread function and a concomitant loss of ability to describe the image formation process linearly.
Gao et al. developed this further, using the traditional reconstruction approach of SR-SIM to enhance effective resolution further still and maintain linear image formation \cite{gao_noninvasive_2012}.
Here, the authors showed that this Bessel light sheet structured illumination approach provided clearer images of live LLC-PK1 cells in anaphase than wide-field SIM, despite the superior theoretical resolution of the latter.
In the final development of Bessel light sheet SIM, Chen, Legant \& Wang et al. used a coherent superposition of Bessel beams to create a so-called `lattice' light sheet, featuring improved modulation contrast and a higher duty-cycle \cite{chen_lattice_2014}.
However, one drawback of all these approaches is that only one pattern orientation is used, resulting in an anisotropic resolution enhancement.
This is further discussed in \prettyref{sec:light_sheet}.

The prototypical example of combining SIM with confocal microscopy is image scanning microscopy, discussed in \prettyref{sec:ism}.
One of the main drawbacks of this approach is the need to acquire an entire image for each scan position of the confocal spot, meaning that hundreds or thousands of images must be recorded to reconstruct one enhanced resolution image.
As such, multifocal structured illumination microscopy (MSIM) was introduced, using multiple spots of illumination in parallel \cite{york_resolution_2012}.
Following this, in was realised independently by multiple research groups that the reconstruction procedure was amenable to an all-optical implementation \cite{de_luca_re-scan_2013,roth_optical_2013,york_instant_2013}.
A further approach, using a 32-segment confocal detector and multichannel photomultiplier tube has also been introduced in Zeiss' Airyscan product \cite{huff_airyscan_2015}.

In all these approaches, the use of a confocal pinhole (either physical or digital) introduces optical sectioning, making these techniques less sensitive to out-of-focus light than traditional SIM.
However, the effective resolution and contrast in non-scattering samples is lower, due to the less effective way in which the illumination brings higher spatial frequencies into the passband.
Compared with traditional confocals, apart from the need for smaller pixel pitches to satisfy the Nyquist-Shannon sampling criterion, there is no drawback to using these SIM-like approaches, with benefits not only in resolution but in signal-to-noise ratio, due to the superconcentration of light \cite{sheppard_interpretation_2016,roth_superconcentration_2016}.

\section{Should image scanning microscopy be considered a form of SIM and what forms of structured illumination could be used other than stripes?}\label{sec:ism}
Image scanning microscopy (ISM), first proposed by Sheppard in 1988 \cite{sheppard_super-resolution_1988} and first demonstrated by M\"uller and Enderlein in 2010 \cite{muller_image_2010}, is experimentally quite different from SIM.
In ISM, a tightly-focussed illumination spot is scanned over the sample, with a camera image of the emission recorded at every scan position.
For each position in the spot image, the sample location most likely to have produced that intensity is not, as one might assume, at the location corresponding to the image position, but instead is located \(\beta / 2\) between the illumination scan position and image position, where \(\beta = \lambda_\textrm{emission} \big/ \lambda_\textrm{illumination}\) is the Stokes' shift factor (i.e. for no Stokes' shift, the most-likely location is halfway between).

In ISM processing, every pixel in each image is relocated in this way, with the corresponding intensity being added to an upsampled pixel grid.
At first, this seems very different from the SIM discussed so far.
However, the idea that non-uniform illumination can provide higher resolution information is still the same, albeit with a very different form of structured illumination.

\begin{figure*}
	\centering
	\includegraphics[width=\textwidth]{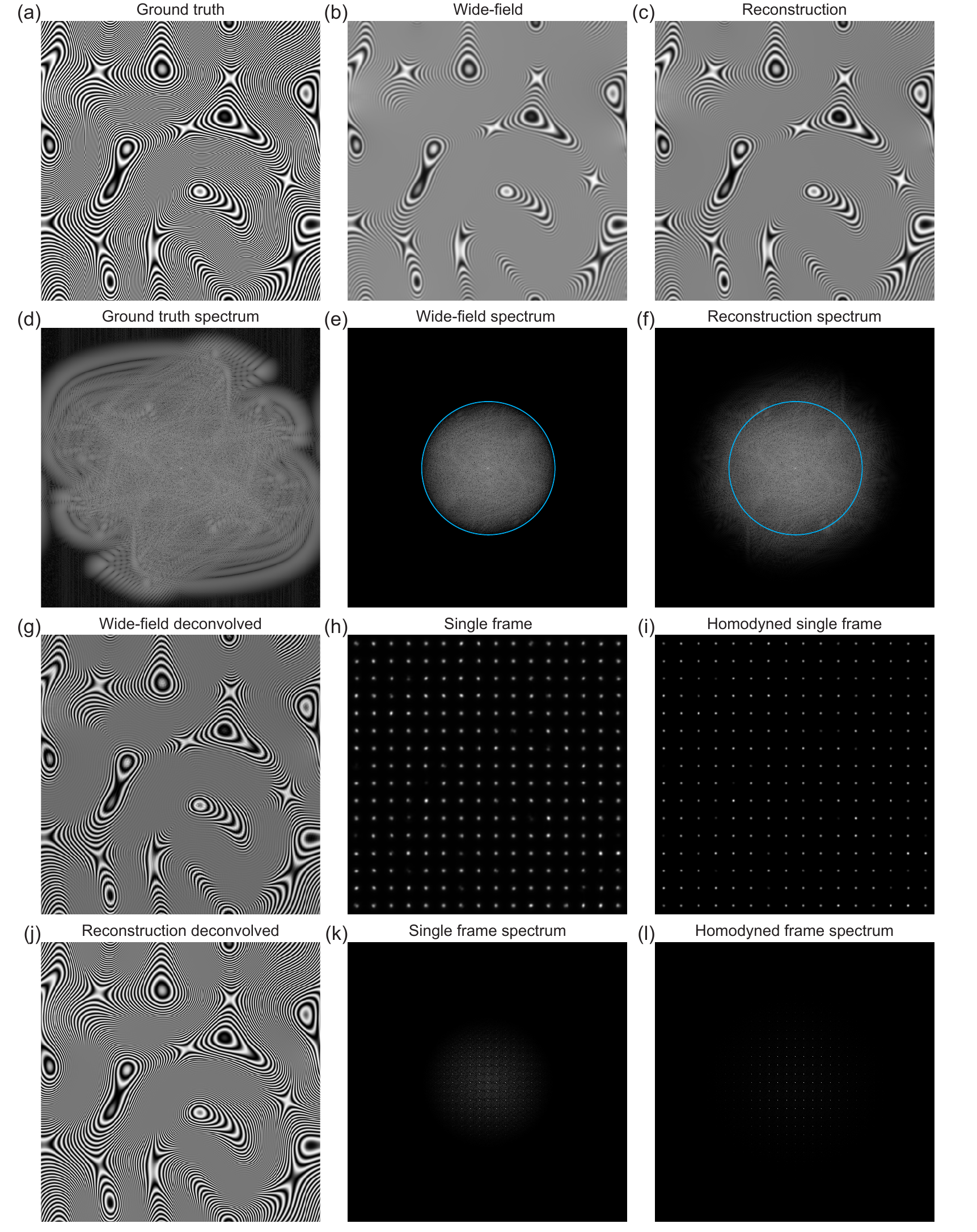}
	\caption{\label{fig:ism_processing}
		The real-space method applied to (parallelised) ISM data (MSIM).
		(a) Ground truth sample structure.
		(b) Wide-field image obtained with uniform illumination.
		(c) Complete SIM reconstruction.
		(d,e,f) Logarithmically-scaled Fourier spectra of (a,b,c).
		(g) \new{Linearly} deconvolved wide-field image.
		(h) Single frame of acquisition.
		(i) Frame from (h) multiplied by local oscillator.
		(j) \new{Linearly} deconvolved SIM reconstruction.
		(k,l) Logarithmically-scaled Fourier spectrum of (h,i).
		Cyan cicles in (e,f) correspond to the bandlimit of the wide-field data.
	}
\end{figure*}

Indeed, the real-space reconstruction approach introduced in \prettyref{sec:fourier} can be extended to process ISM data as well\new{, albeit not in a way that uses photon reassignment}.
For the purposes of visualisation and computational speed, \prettyref{fig:ism_processing} uses multiple focussed illumination spots in parallel, as produced by MSIM \cite{york_resolution_2012}.
As before, each raw image (\prettyref{fig:ism_processing}h) is multiplied by a pattern that mimics the spatial period and phase of the illumination, but is spectrally flat.
These multiplied images (\prettyref{fig:ism_processing}i) are then summed to produce the final reconstruction (\prettyref{fig:ism_processing}c).
As shown by a comparison of the spectra in \prettyref{fig:ism_processing}e and f, the processed image contains information at higher spatial frequencies than in the simple sum of the raw data.

\begin{figure*}
	\centering
	\includegraphics[width=\textwidth]{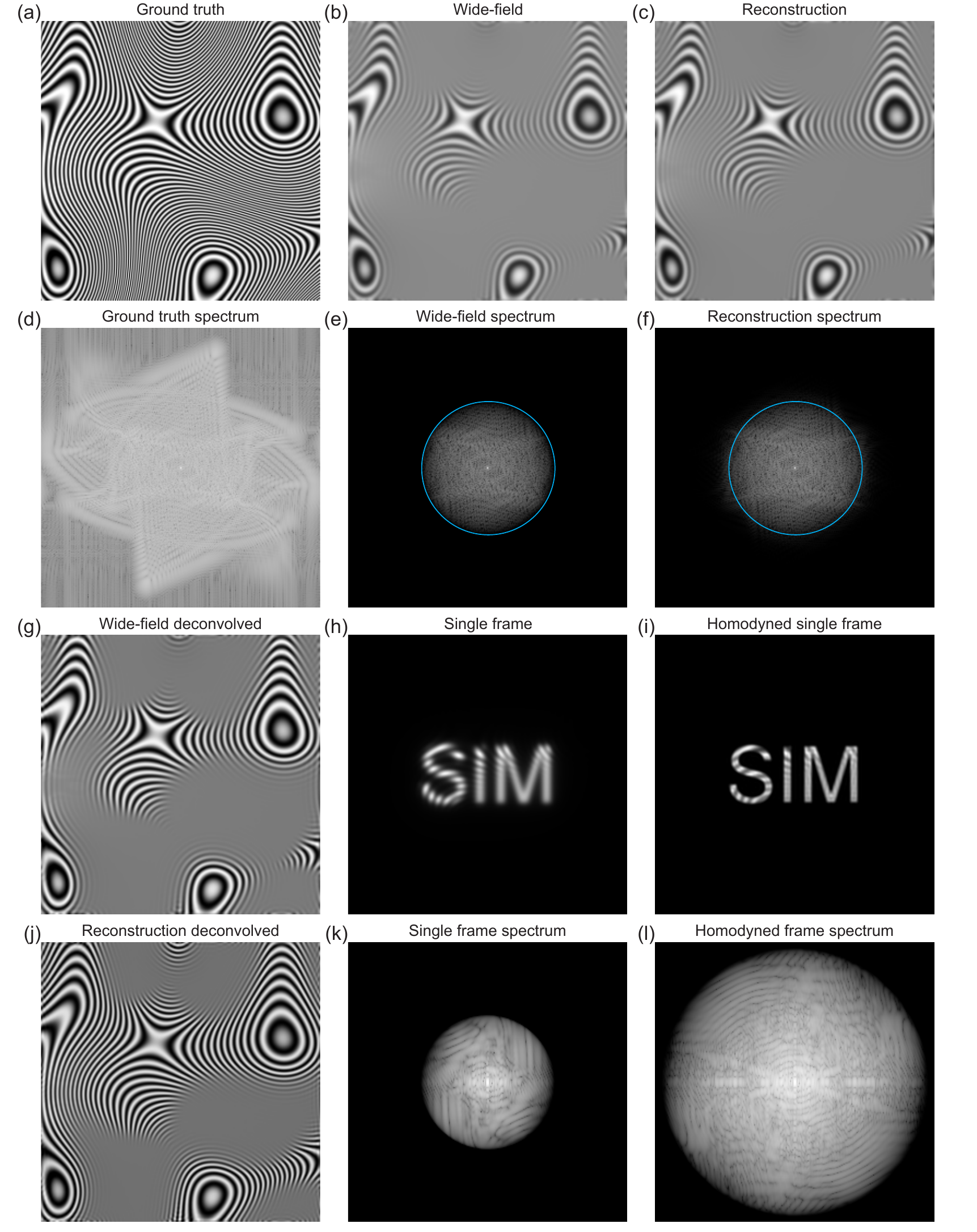}
	\caption{\label{fig:arbitrary_processing}
		The real-space method applied to an ISM-like acquisition where, instead of a focussed spot, a projection of the text `SIM' is used for illumination.
		(a) Ground truth sample structure.
		(b) Wide-field image obtained with uniform illumination.
		(c) Complete SIM reconstruction.
		(d,e,f) Logarithmically-scaled Fourier spectra of (a,b,c).
		(g) \new{Linearly} deconvolved wide-field image.
		(h) Single frame of acquisition.
		(i) Frame from (h) multiplied by local oscillator.
		(j) \new{Linearly} deconvolved SIM reconstruction.
		(k,l) Logarithmically-scaled Fourier spectrum of (h,i).
		Cyan cicles in (e,f) correspond to the bandlimit of the wide-field data.
	}
\end{figure*}

Furthermore, the real-space reconstruction approach can be modified to handle quasiarbitrary illumination patterns.
As an example of this, \prettyref{fig:arbitrary_processing} shows a workflow for using a projection of the text `SIM' as an illumination pattern.
\new{\prettyref{svid:sim_ism} shows the steps in the acquisition and reconstruction as data is collected.}
While not necessarily spectrally optimal, a comparison of the spectra of the raw sum and remultiplied sum, along with \new{linearly} deconvolved images, shows that higher resolution information has successfully been extracted.
An alternative approach in which the illumination pattern is not known, called `blind SIM', has also been investigated \cite{mudry_structured_2012,ayuk_structured_2013,jost_optical_2015,yeh_structured_2017}.

\section{How does sparse illumination compare to dense illumination in linear and non-linear SIM?}\label{sec:sparsity}
As discussed in \prettyref{sec:contrast}, SIM requires the illumination pattern to have good contrast in order to work well.
For thicker samples, where much out-of-focus light contaminates the image of the plane of interest, sparse illumination is helpful.
However, for maximum performance at high spatial frequencies, a dense illumination is required.
In addition, a greater number of phase shifts must be used.

This can be understood by considering real space and the Fourier domain as reciprocal spaces.
Hence, sparse illumination results in a dense Fourier spectrum and vice versa.
A dense Fourier spectrum means that more components must be unmixed, and so a greater number of phase shifts must be used.

\begin{figure*}
	\centering
	\includegraphics[width=\textwidth]{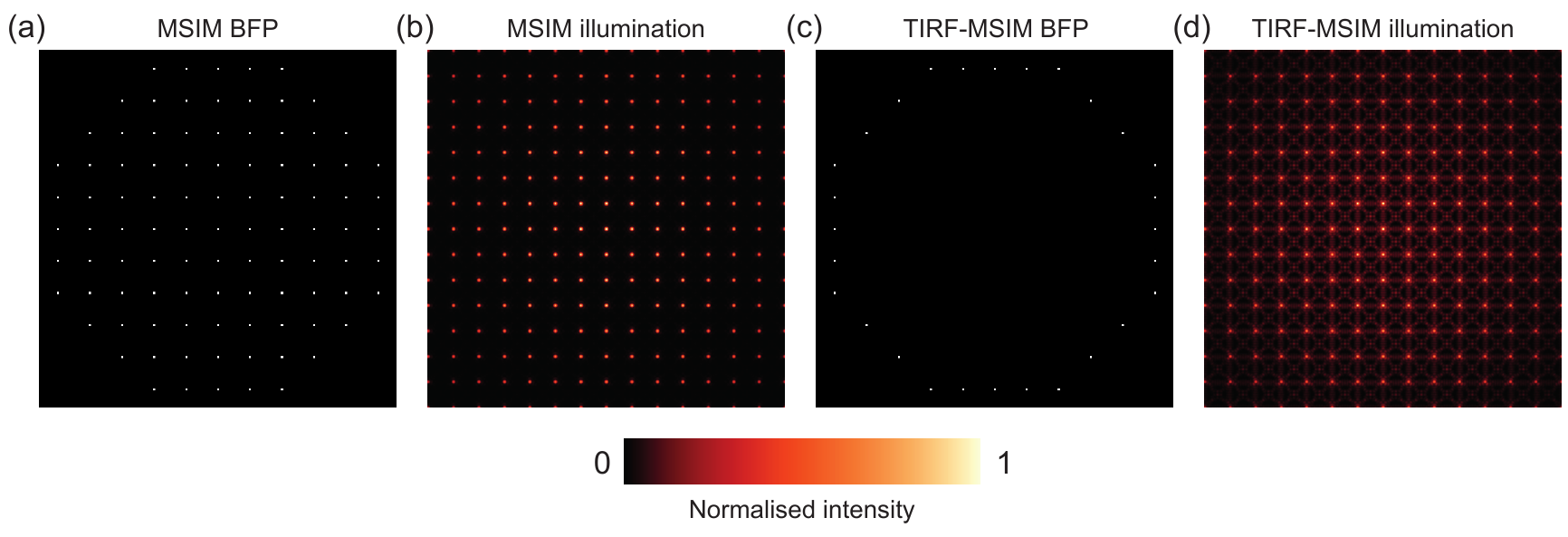}
	\caption{\label{fig:tirf-msim}
		Simulated back focal planes (a,c) and illumination patterns (b,d) for MSIM (a,b) and TIRF-MSIM (c,d).
		A non-linear colourmap is used to enhance the visibility of the undesired interference patterns between spots in (d).
	}
\end{figure*}

Furthermore, dense illumination is more readily combined with total internal reflection fluorescence microscopy.
To see this, consider an MSIM-like illumination pattern, where the back focal plane of the objective has the illumination shown in \prettyref{fig:tirf-msim}a\new{, with circular polarisation}.
This produces an illumination in the focal plane as shown in \prettyref{fig:tirf-msim}b.
If we wished for the illumination to be confined to the TIRF-regime, then we must ensure that only the outer annulus of the back focal plane is illuminated, as shown in \prettyref{fig:tirf-msim}c.
This results in the illumination shown in \prettyref{fig:tirf-msim}d.
Here, the interference in the pattern causes unwanted illumination, leading to excess photodamage and noise. 

Recently, Ingerman et al. considered a related question specifically for dense illuminations: is it better to use two-dimensional patterns or sequentially rotated one-dimensional patterns for linear and non-linear SIM \cite{ingerman_signal_2019}?
Perhaps surprisingly, they found that two-dimensional patterns were better for \new{high-resolutions} in non-linear SIM while one-dimensional patterns were superior for linear SIM, although the performance advantage of 2D patterns disappeared when non-photoswitchable background was introduced.
\new{In addition, for the non-linear case, they found that two-dimensional patterns require almost two-fold more exposure to off-switching light as one-dimensional patterns, for the same resolution enhancement.
This suggests that one-dimensional methods are superior if photodamage from off-switching light is present.}

\new{These considerations began more than a decade before appearing in print, but publication was delayed by the deteriorating health and untimely death of Mats Gustafsson in 2011.
The version initially submitted in 2008 was revived and modified by Ingerman, London and Heintzmann as a tribute, being finally published in 2019.}

\section{Can SIM be used to improve the resolution of (Rayleigh scattering) transmission microscopy?}\label{sec:scattering}
No.
As noted in \prettyref{sec:abbe}, the resolution limit for oblique illumination is already twice as good as that for head-on illumination.
A structured illumination setup can provide such oblique illumination, but there is no further resolution to be gained.
For further information, see the excellent article by Wicker \& Heintzmann \cite{wicker_resolving_2014}.

\section{Can we generate `true' super-resolution images from simple instruments enhanced with machine-learning-based algorithms?}\label{sec:machine_learning}
Machine learning has proven to be a powerful tool for diverse image processing applications, such as image segmentation, denoising, deconvolution etc. \cite{arganda-carreras_trainable_2017,schmidt_cell_2018,boulanger_patch-based_2010,guo_rapid_2020}.
However, care should be taken to distinguish between approaches that try to \emph{extract} information from an image and those that try to \emph{add} information to an image.
With good training, it has been possible to develop multiple deep-learning approaches that convert an image, or set of images (such as a SIM acquisition sequence), into a higher resolution image \cite{ling_fast_2020,jin_deep_2020,qiao_evaluation_2021,shah_deep-learning_2021}.
While processing a set of images may fall into the `extract' category, processing a single image in such a way must necessarily fall into the `add' category.
As such, machine-learning will never be able to truly replace, rather than augment, a super-resolution technique.
Nevertheless, such single-image approaches may be able to improve the effective resolution in a manner similar to deconvolution.

\section{Can research-grade super-resolution (SIM) microscopes be built cost-efficiently?}\label{sec:cost}
All structured illumination microscopes are relatively cheap --- the expense lies in the time required to make them work well!
A basic system can be constructed using a rotation mount, piezo translation mount, diffraction grating, polariser and a standard laser, plus some lenses and mirrors (as in the very first SIM publications \cite{heintzmann_laterally_1999,gustafsson_surpassing_2000,gustafsson_three-dimensional_2008}).
However, such a system will be too slow for live-cell imaging and can severely limit the throughput of fixed-sample imaging \cite{kner_super-resolution_2009,shao_super-resolution_2011}.
Constructing a SIM system fast enough for live-cell imaging requires a way of generating the sinusoidal illumination pattern without physical rotation of a grating.
In general, two methods are common: spatial light modulators (SLMs) and interferometric systems.

Spatial light modulators can, to a first approximation, be thought of as reprogrammable diffraction gratings and used in a similar way.
However, instead of shifting and rotating the SLM, the pattern displayed by the SLM is instead updated to match.
This arrangement maintains the common-path layout of grating-based systems, minimising the sensitivity to air currents and temperature fluctuations.
However, fast-switching ferroelectric SLMs have a \new{very low} diffraction efficiency (often \(< 5\%\)) and so require powerful laser sources to be used.
\new{Fast-switching nematic SLMs, with much higher diffraction efficiencies, are now commercially available, albeit at high cost and not yet capable of the frame rates of ferroelectric SLMs.}

Interferometric systems normally feature some kind of Michelson/Twyman--Green or Mach--Zehnder interferometer, where pattern orientation is changed by changing optical paths and phase steps are accomplished by changing the length of one arm of the interferometer \cite{best_structured_2011,liu_three-dimensional_2019}.
These approaches are much more photon-efficient than SLM-based systems and, as the pattern period changes with wavelength, rather than being fixed, can be used to ensure the same effective illumination NA over all wavelengths in use.
This is the approach taken by the commercial OMX Blaze system \cite{dougherty_method_2013}.
However, such interferometers are extremely sensitive to air currents and temperature fluctuations, as any phase shift in one arm of the interferometer directly couples to a phase shift in the illumination pattern.
In an attempt to circumvent this issue, fibre-based interferometeric SIM systems have recently been developed \cite{hinsdale_high-speed_2021,pospisil_highly_2021}.

Overall, SLM-based systems are by far the most common, in both research-grade and commercial instruments.
In particular, Forth Dimension Displays have cemented themselves as the preferred \new{supplier} of ferroelectric SLMs for the SIM community, due to their high pixel count and rapid switching speed.
Nevertheless, successful operation of such an SLM is tricky for a newcomer, due to effects from the binary nature of the pixels, interpixel vias, duty-cycle requirements and low photon-efficiency.
As such, there has recently been much interest in the development of systems that use a different type of SLM, namely digital micromirror devices (DMDs) \cite{sandmeyer_dmd-based_2019,brown_multicolor_2020}.
Found in many projectors, these devices rely on tilting tiny micromirrors to direct light in one direction or another, acting as a binary amplitude display.
While these provide high photon-efficiencies, their tilted-mirror nature means that they act akin to blazed gratings, with a strongly wavelength-dependent distribution of power into different diffraction orders.
As such, their use is currently restricted to those labs interested in researching how best to make use of DMDs for SIM, rather than following the tried-and-tested approach of ferroelectric SLMs and powerful lasers.

\section{How can information about single-molecule detection be best combined with the knowledge of the illumination structure?}\label{sec:smlm}
This is an area of ongoing research, prompted by the development of MINFLUX \cite{balzarotti_nanometer_2017}.
In MINFLUX, an intensity minimum at the centre of a doughnut beam is used to localise single fluorophores by iteratively updating the doughnut position and minimising the photon flux.
While this approach can provide excellent localisation precisions (\(< \SI{1}{nm}\)), it is inherently slow as it only probes one molecule at a time.
Since the initial publication of the MINFLUX method, there have been a number of publications which combine full-field fringe-like structured illumination with the MINFLUX concept to provide high-precision localisations at faster rates \cite{reymond_simple_2019,gu_molecular_2019,cnossen_localization_2020,jouchet_nanometric_2021}.
Most recently, Schmidt et al. have shown that these methods only use half the localisation information encoded by the structured illumination and demonstrate a further method that, unlike previous approaches, achieves the Cram\'er--Rao lower bound \cite{schmidt_camera-based_2021}.

\section{Can the axial resolution of a SIM microscope be enhanced beyond a two-fold improvement?}\label{sec:axial}
Enhancing the axial resolution of a SIM system requires accessing the sample from the other side, either to introduce extra wavevectors for illumination, or to capture extra wavevectors of fluorescence.
Perhaps the most impressive SIM microscope yet built, the I\textsuperscript{5}S of Shao et al., does both \cite{shao_i5s:_2008}.
By using a second opposing high-numerical-aperture objective, a further three beams of illumination are introduced.
The same objective is also used to capture fluorescence, with the two objectives being aligned to such a precision that the fluorescence captured by both can be made to interfere via a beamsplitter, despite the \textasciitilde\SI{1}{\micro m} coherence length of the light.
This provides an almost-isotropic, better-than-\SI{100}{nm} resolution in all three dimensions, providing volumes of objects with no preferred viewing direction.

Despite this impressive performance, I\textsuperscript{5}S is not widely used, due to the challenging alignment and stability requirements for its operation.
In addition, the very short coherence length of fluorescence emission (\textasciitilde\SI{1}{\micro m}) means that sample-induced phase differences restrict the technique to imaging thin single cells.
Recently, Manton et al. proposed that a lower-numerical-aperture, long-working-distance dipping objective could be used to introduce a single counterpropagating beam and provide much of the axial resolution enhancement of I\textsuperscript{5}S in a much simpler setup \cite{manton_concepts_2020}.
While such a system is theoretically capable of providing 3D resolutions better than \SI{125}{nm}, the transfer function at high spatial frequencies is lower than that of an I\textsuperscript{5}S microscope without interferometric detection.
As such, we expect that six-beam structured illumination provided by two opposing high-numerical-aperture objectives will be the future method of choice for extended-resolution imaging when isotropic resolution is key.

\section{Can light sheet microscopy and SIM, with multiple pattern orientations, be properly combined?}\label{sec:light_sheet}
As noted in \prettyref{sec:contrast}, light sheet and SIM appear to be a perfect match: light sheet microscopy is limited in resolution due to the relatively low numerical aperture objectives used, but generates very little out-of-focus light --- the bane of SIM.
However, traditional light sheet SIM (LSFM-SIM) suffers from the relatively coarse illumination pattern that can be provided by the excitation objective in LSFM systems and the fact that only one orientation of the pattern is present.
As such, while improvements in axial resolution can be made, very little useful enhancement of lateral resolution is possible.

In an attempt to counter these issues, Manton \& Rees proposed a LSFM-SIM system featuring three mutually orthogonal objectives \cite{manton_trispim:_2016}.
However, the use of 0.8 NA objectives, necessary for reasons of steric hindrance, meant that the maximum possible lateral resolution would still only be around \SI{230}{nm} (i.e. approximately the theoretical resolution of a 1.1 NA water-dipping objective as used in some other light sheet microscope designs).
While they also suggested the use of interferometric excitation through two objectives to improve the resolution to \SI{120}{nm}, the experimental complexity of such a system means that it is unlikely to ever be realised in practice.

In 2017, Chang et al. demonstrated a three-objective system featuring interferometric excitation through two objectives and counterpropagating light sheets \new{\cite{chang_csilsfm_2017}}.
While the use of counterpropagating sheets produces a fine fringe period, issues with steric hindrance limited the orientations over which patterns could be produced, precluding a reconstruction with isotropic resolution.
In addition, the large distance between arms of the interferometer mean that the system is sensitive to temperature-induced phase shifts, as discussed in \prettyref{sec:cost}, \new{and sample mounting is made more difficult}.

\begin{figure*}
	\centering
	\includegraphics[width=\textwidth]{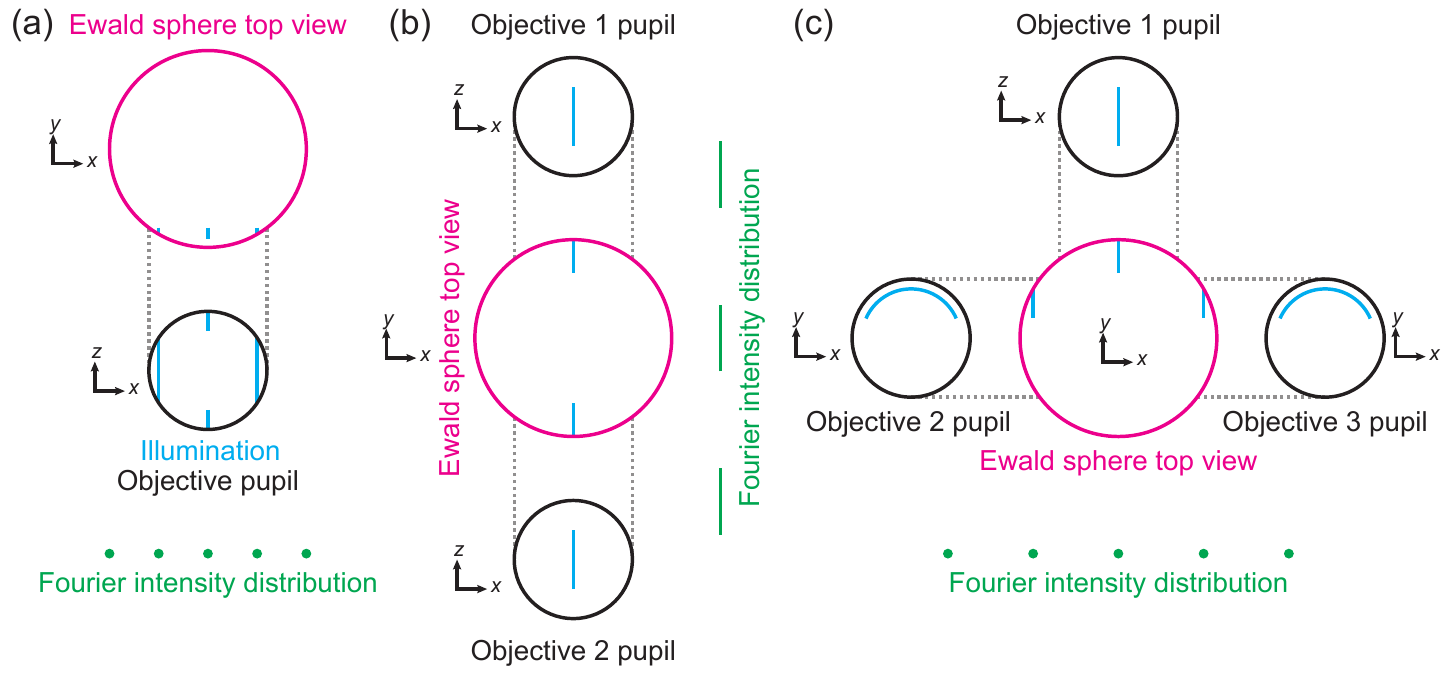}
	\caption{\label{fig:lsfm_ewald}
		Pupils and Ewald spheres for light sheet structured illumination microscopy approaches.
		(a) Single pupil and illumination pattern for lattice light sheet microscopy.
		As desired, the Fourier intensity distribution consists of harmonic functions laterally.
		(b) Opposed pupils using counterpropagating light sheets for SIM.
		Here, the components of the Fourier intensity distribution are stretched out along the pattern direction, precluding conventional reconstructions in which harmonic components are assumed.
		(c) A potential method for producing lattice-like illumination patterns at spatial frequencies beyond that achievable with one objective.
		Here, C-shaped illumination profiles in the auxilliary objectives are used to introduce the off-centre Ewald sphere components, ensuring a harmonic Fourier intensity distribution.
	}
\end{figure*}

Furthermore, there is a theoretical issue with using interferometric excitation through multiple objectives that has so far been unaddressed.
In the traditional SIM approach, we consider the illumination to be described by a \new{discrete} sum of purely harmonic functions (i.e. functions which consist of a single spatial frequency).
As shown in \prettyref{fig:lsfm_ewald}a, this condition is satisfied for lattice light sheet microscopy \cite{chen_lattice_2014} as the pupil pattern consists solely of vertical lines, which project orthographically onto the Ewald sphere (i.e. they still look like straight lines when viewed from above, with the separation between lines being constant --- the distribution is a \new{discrete} sum of harmonic functions along the pattern direction).
However, now consider a system in which a single vertical line in the pupil is used to produce a light sheet with one objective, with another identical pupil and objective opposing (see \prettyref{fig:lsfm_ewald}b).
Now, when the Ewald sphere is viewed from above, the distribution is not a \new{discrete} sum of harmonic functions along the pattern direction.
While the limitations on numerical aperture provided by use of a single objective can be circumvented by using C-shaped illumination profiles in ancillary objective pupils (\prettyref{fig:lsfm_ewald}c), it appears that the use of counterpropagating sheets is not allowed, at least not by standard reconstruction approaches.

\section{How quickly can SIM image a sample?}\label{sec:speed}
Fundamentally, the speed of SIM is restricted by the requirement that enough photons are detected for the signal-to-noise ratio to permit a high-quality reconstruction.
So far, the fastest SIM acquisitions demonstrated were acquired at \SI{798}{Hz} and \SI{714}{Hz}, corresponding to image frame rates of \SI{88.7}{Hz} and \SI{79}{Hz} \cite{guo_visualizing_2018,song_fast_2016}.
Given that even a single fluorophore can emit at rates in excess of 100 million photons per second, it appears that current speed limits are more practical than fundamental.

In an attempt to reduce the acquisition time for volumetric data by reducing the number of stage movements that must be completed, SIM has been combined with multifocal microscopy \cite{prabhat_simultaneous_2004}.
Here, images from multiple depths within the sample are recorded at the same time.
While the initial demonstration in 2017, by Abrahamsson et al., used an aberration-free multifocus scheme, the limited speed of the commercial SIM system used prevented fast acquisitions \cite{abrahamsson_multifocus_2017}.
A more recent demonstration, in 2020, of high-speed acquisition used an aberration-inducing multifocus prism, limiting the depth over which images could be acquired in parallel \cite{descloux_high-speed_2020}.

While these experiments demonstrate an impressive effort in pushing hardware to its limits, other effort has been expended on improving the reconstruction process.
To enable live visualisation of a sample with extended resolution, Markwirth et al. developed a software system that can provide GPU-accelerated reconstructions in almost real-time \cite{markwirth_video-rate_2019}.
Many groups, along with Zeiss in their Elyra 7, have adopted `interleaved reconstruction', in which rolling sets of three orientations are used to produce reconstructions, rather than only producing one image from each group of three-orientation data \cite{ma_structured_2018}.
While this produces smoother videos, it is a form of interpolation and does not truly increase speed \cite{boualam_method_2021}.

A number of attempts have been made at reducing the number of raw images required for a successful reconstruction, down from nine.
Heintzmann was the first to propose that a smaller set of images could be sufficient, identifying that using as few as four images might be possible \cite{heintzmann_saturated_2003}.
Lal et al. have proposed such a reconstruction using just four images, but the illumination patterns used do not average out to a uniform intensity \cite{lal_frequency_2018}.
This means that, for time-lapse imaging, the illumination pattern would `burn' into the sample via spatially-dependent photobleaching.
Orieux et al. have developed a Bayesian approach to SIM reconstruction and used this to develop a method that uses four images and has no specific constraints on the illumination \cite{orieux_bayesian_2012}.
Furthermore, Str\"ohl and Kaminski investigated the possibility of a further reduction to just three images and developed a maximum likelihood approach to deal with the resulting underdetermined system \cite{strohl_speed_2017}.

\section{Outlook}\label{sec:outlook}
\begin{figure*}
	\centering
	\includegraphics[width=\textwidth]{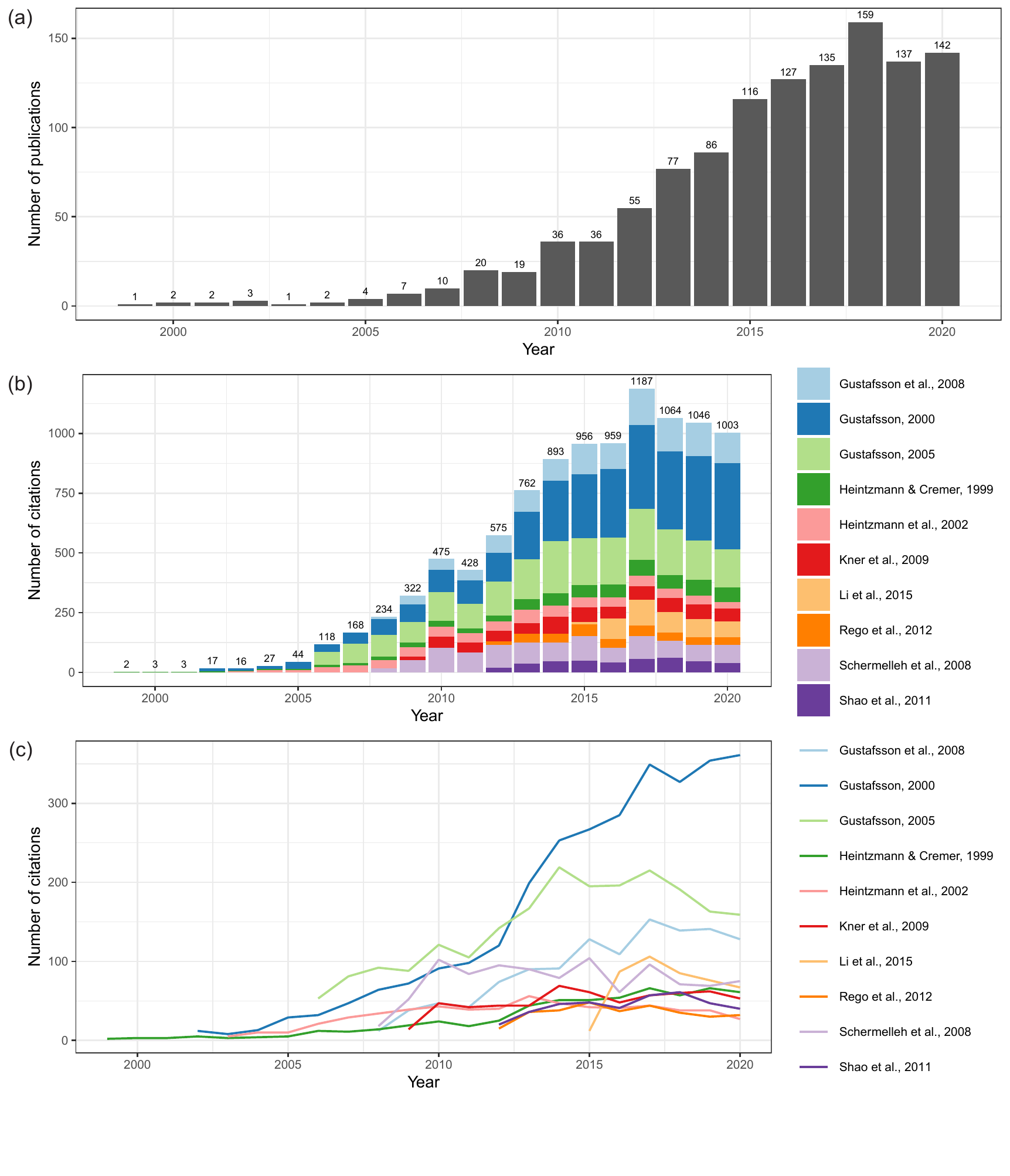}
	\caption{\label{fig:bibliometrics}
		A bibliometric analysis of SIM publications, from inception to the end of 2020.
		(a) shows the number of PubMed-indexed SIM publications per year, peaking at 159 in 2018.
		(b) shows the number of citations to the top 10 most-cited SIM publications, with the number at the top of each bar stack indicating the total number of citations of the cohort that year, peaking at 1187 in 2017.
		(c) shows the same citation information as (b), but now unstacked.
		This view of the data demonstrates that the peak in 2017 is not due to any one individual publication.
	}
\end{figure*}

At the time of writing, SIM is more than 22 years old.
A PubMed analysis, searching for publications with the phrase ``structured illumination'' in their title or abstract suggests that more than 100 SIM-related articles have been published every year since 2015 (\prettyref{fig:bibliometrics}a).
An analysis of the number of citations, as indexed by Google Scholar, of the top 10 most-cited SIM papers shows that the total number of citations of this cohort now exceeds 1000 per year (\prettyref{fig:bibliometrics}b).
Ignoring the unusual peak in 2017/18, it appears that interest in SIM continues to increase.

Indeed, the estimation of the number of SIM-related papers is likely to be a significant underestimate --- SIM as a technique has now matured to the point where it is no longer a `shiny new toy' that must be mentioned in the title or abstract.
Nowadays, it suffices to just mention SIM in the main text alongside well-respected techniques such as confocal microscopy.

Perhaps the most important development in the past few years is that of all-optical versions of ISM, providing extended-resolution imaging at high frame rates and without the need for computational reconstruction \cite{de_luca_re-scan_2013,roth_optical_2013,york_instant_2013,azuma_super-resolution_2015}.
These instruments present the user with a system that effectively works as a familiar confocal or spinning disk microscope, albeit one with double the resolving power.
Furthermore, by avoiding computational reconstruction, these approaches avoid the reconstruction artefacts that can often plague SIM \cite{demmerle_strategic_2017}.
However, in the absence of background fluorescence, at high spatial frequencies traditional SIM still provides superior contrast and is more compatible with TIRF imaging.
As such, further investigation into which techniques provide better performance in different samples and imaging scenarios will be welcome, as would an all-optical SIM reconstruction approach.

For longer-term time-lapse imaging, light sheet fluorescence microscopy (LSFM) is now the technique of choice for minimal sample damage where fluorescence contrast is required.
While combinations of LSFM and SIM have already been demonstrated (see \prettyref{sec:contrast} and \prettyref{sec:light_sheet}), an instrument that achieves maximal performance of both components has yet to be demonstrated.

Further enhancements to the capabilities of SIM for time-lapse imaging are being provided by enhanced reconstruction routines that use various forms of regularisation to successfully reconstruct images with lower signal levels \cite{boulanger_nonsmooth_2018,huang_fast_2018}.
In combination with brighter fluorophores, further development of these techniques and other denoising procedures will help provide gentle, extended-resolution imaging of proteins of interest at endogenous expression levels.

As well as enhancing the resolution of images and movies, SIM has also been demonstrated to enhance the resolution of biophysical microscopy methods.
While fluorescence lifetime imaging microscopy (FLIM) has been combined with OS-SIM \cite{hinsdale_optically_2017} and used corretively with SR-SIM \cite{gorlitz_mapping_2017}, so far no true combination of FLIM and SIM has been achieved.
In 2019, Zhanghao et al. reported their method of polarised structured illumination microscopy (pSIM), providing extended-resolution imaging of dipoles in biopolymers such as cytoskeletal networks and \(\lambda\)-DNA \cite{zhanghao_super-resolution_2019}.
Most recently, in 2019 Colin-York et al. have combined traction force microscopy (TFM) with 3D-SIM \cite{colin-york_spatiotemporally_2019}, while in 2021 Barbieri et al. combined TFM with TIRF-SIM \cite{barbieri_two-dimensional_2021}.

Altogether, it seems that there is still much room for further work in SIM, both in terms of technique development and applications.
It will be interesting to survey the field again in 20 years' time and see how many of the outstanding challenges have been solved and which unanticipated developments arose.
If the past 20 years are any guide, the answers will be surprising\dots

\section*{Acknowledgements}
The author is grateful to J\'er\^ome Boulanger, Reto Fiolka, Rainer Heintzmann, Chris Rowlands, Florian Str\"{o}hl and Andrew York for various SIM-related discussions/arguments over the years, and thanks J\'er\^ome Boulanger for the ground-truth image used in Figures 3, 4 and 5.
The author acknowledges support from Fitzwilliam College, Cambridge, through a Research Fellowship.

\printbibliography

@inproceedings{lin_structured_2021,
	title = {Structured illumination microscopy with direct wavefront sensing},
	volume = {11652},
	url = {https://www.spiedigitallibrary.org/conference-proceedings-of-spie/11652/116520A/Structured-illumination-microscopy-with-direct-wavefront-sensing/10.1117/12.2582806.full},
	doi = {10.1117/12.2582806},
	eventtitle = {Adaptive Optics and Wavefront Control for Biological Systems {VII}},
	pages = {10--13},
	booktitle = {Adaptive Optics and Wavefront Control for Biological Systems {VII}},
	publisher = {{SPIE}},
	author = {Lin, Ruizhe and Kner, Peter A.},
	urldate = {2021-11-15},
	date = {2021-03-05},
}

@article{hell_toward_2003,
	title = {Toward fluorescence nanoscopy},
	volume = {21},
	rights = {2003 Nature Publishing Group},
	issn = {1546-1696},
	url = {https://www.nature.com/articles/nbt895},
	doi = {10.1038/nbt895},
	pages = {1347--1355},
	number = {11},
	journaltitle = {Nature Biotechnology},
	shortjournal = {Nat Biotechnol},
	author = {Hell, Stefan W.},
	urldate = {2021-11-15},
	date = {2003-11},
	langid = {english},
	note = {Bandiera\_abtest: a
Cg\_type: Nature Research Journals
Number: 11
Primary\_atype: Reviews
Publisher: Nature Publishing Group},
}

@article{schwentker_wide-field_2007,
	title = {Wide-field subdiffraction {RESOLFT} microscopy using fluorescent protein photoswitching},
	volume = {70},
	issn = {1097-0029},
	url = {https://onlinelibrary.wiley.com/doi/abs/10.1002/jemt.20443},
	doi = {10.1002/jemt.20443},
	pages = {269--280},
	number = {3},
	journaltitle = {Microscopy Research and Technique},
	author = {Schwentker, Miriam A. and Bock, Hannes and Hofmann, Michael and Jakobs, Stefan and Bewersdorf, Jörg and Eggeling, Christian and Hell, Stefan W.},
	urldate = {2021-11-15},
	date = {2007},
	langid = {english},
	note = {\_eprint: https://onlinelibrary.wiley.com/doi/pdf/10.1002/jemt.20443},
}

@inproceedings{hirvonen_structured_2008,
	title = {Structured illumination microscopy using photoswitchable fluorescent proteins},
	volume = {6861},
	url = {https://www.spiedigitallibrary.org/conference-proceedings-of-spie/6861/68610L/Structured-illumination-microscopy-using-photoswitchable-fluorescent-proteins/10.1117/12.763021.full},
	doi = {10.1117/12.763021},
	eventtitle = {Three-Dimensional and Multidimensional Microscopy: Image Acquisition and Processing {XV}},
	pages = {133--140},
	booktitle = {Three-Dimensional and Multidimensional Microscopy: Image Acquisition and Processing {XV}},
	publisher = {{SPIE}},
	author = {Hirvonen, Liisa and Mandula, Ondrej and Wicker, Kai and Heintzmann, Rainer},
	urldate = {2021-11-15},
	date = {2008-02-12},
}

@article{bachl_experiments_1967,
	title = {Experiments on Superresolution Imaging of a Reduced Object Field*},
	volume = {57},
	rights = {\&\#169; 1967 Optical Society of America},
	url = {https://www.osapublishing.org/josa/abstract.cfm?uri=josa-57-2-163},
	doi = {10.1364/JOSA.57.000163},
	pages = {163--169},
	number = {2},
	journaltitle = {{JOSA}},
	shortjournal = {J. Opt. Soc. Am., {JOSA}},
	author = {Bachl, A. and Lukosz, W.},
	urldate = {2021-11-15},
	date = {1967-02-01},
	note = {Publisher: Optical Society of America},
}

@article{lukosz_optical_1967,
	title = {Optical Systems with Resolving Powers Exceeding the Classical Limit. {II}},
	volume = {57},
	rights = {\&\#169; 1967 Optical Society of America},
	url = {https://www.osapublishing.org/josa/abstract.cfm?uri=josa-57-7-932},
	doi = {10.1364/JOSA.57.000932},
	pages = {932--941},
	number = {7},
	journaltitle = {{JOSA}},
	shortjournal = {J. Opt. Soc. Am., {JOSA}},
	author = {Lukosz, W.},
	urldate = {2021-11-15},
	date = {1967-07-01},
	note = {Publisher: Optical Society of America},
}

@article{lukosz_optical_1966,
	title = {Optical Systems with Resolving Powers Exceeding the Classical Limit*},
	volume = {56},
	rights = {\&\#169; 1966 Optical Society of America},
	url = {https://www.osapublishing.org/josa/abstract.cfm?uri=josa-56-11-1463},
	doi = {10.1364/JOSA.56.001463},
	pages = {1463--1471},
	number = {11},
	journaltitle = {{JOSA}},
	shortjournal = {J. Opt. Soc. Am., {JOSA}},
	author = {Lukosz, W.},
	urldate = {2021-11-15},
	date = {1966-11-01},
	note = {Publisher: Optical Society of America},
}

@article{stemmer_widefield_2008,
	title = {Widefield fluorescence microscopy with extended resolution},
	volume = {130},
	issn = {1432-119X},
	url = {https://doi.org/10.1007/s00418-008-0506-8},
	doi = {10.1007/s00418-008-0506-8},
	pages = {807},
	number = {5},
	journaltitle = {Histochemistry and Cell Biology},
	shortjournal = {Histochem Cell Biol},
	author = {Stemmer, Andreas and Beck, Markus and Fiolka, Reto},
	urldate = {2021-11-15},
	date = {2008-09-23},
	langid = {english},
}

@article{chung_extended_2006,
	title = {Extended resolution wide-field optical imaging: objective-launched standing-wave total internal reflection fluorescence microscopy},
	volume = {31},
	rights = {\&\#169; 2006 Optical Society of America},
	issn = {1539-4794},
	url = {https://www.osapublishing.org/ol/abstract.cfm?uri=ol-31-7-945},
	doi = {10.1364/OL.31.000945},
	shorttitle = {Extended resolution wide-field optical imaging},
	pages = {945--947},
	number = {7},
	journaltitle = {Optics Letters},
	shortjournal = {Opt. Lett., {OL}},
	author = {Chung, Euiheon and Kim, Daekeun and So, Peter T.},
	urldate = {2021-11-15},
	date = {2006-04-01},
	note = {Publisher: Optical Society of America},
}

@article{fiolka_structured_2008,
	title = {Structured illumination in total internal reflection fluorescence microscopy using a spatial light modulator},
	volume = {33},
	rights = {\&\#169; 2008 Optical Society of America},
	issn = {1539-4794},
	url = {https://www.osapublishing.org/ol/abstract.cfm?uri=ol-33-14-1629},
	doi = {10.1364/OL.33.001629},
	pages = {1629--1631},
	number = {14},
	journaltitle = {Optics Letters},
	shortjournal = {Opt. Lett., {OL}},
	author = {Fiolka, Reto and Beck, Markus and Stemmer, Andreas},
	urldate = {2021-08-10},
	date = {2008-07-15},
	note = {Publisher: Optical Society of America},
}

@article{gustafsson_extended_1999,
	title = {Extended resolution fluorescence microscopy},
	volume = {9},
	issn = {0959-440X},
	url = {https://www.sciencedirect.com/science/article/pii/S0959440X99000160},
	doi = {10.1016/S0959-440X(99)00016-0},
	pages = {627--628},
	number = {5},
	journaltitle = {Current Opinion in Structural Biology},
	shortjournal = {Current Opinion in Structural Biology},
	author = {Gustafsson, Mats {GL}},
	urldate = {2021-08-10},
	date = {1999-10-01},
	langid = {english},
}

@article{sheppard_fundamentals_2007,
	title = {Fundamentals of superresolution},
	volume = {38},
	issn = {0968-4328},
	url = {https://www.sciencedirect.com/science/article/pii/S0968432806001302},
	doi = {10.1016/j.micron.2006.07.012},
	series = {Special issue on Super-resolution and other Novel Microscopies},
	pages = {165--169},
	number = {2},
	journaltitle = {Micron},
	shortjournal = {Micron},
	author = {Sheppard, Colin J. R.},
	urldate = {2021-07-15},
	date = {2007-02-01},
	langid = {english},
}

@article{heintzmann_answers_2021,
	title = {Answers to fundamental questions in superresolution microscopy},
	volume = {379},
	url = {https://royalsocietypublishing.org/doi/10.1098/rsta.2021.0105},
	doi = {10.1098/rsta.2021.0105},
	pages = {20210105},
	number = {2199},
	journaltitle = {Philosophical Transactions of the Royal Society A: Mathematical, Physical and Engineering Sciences},
	author = {Heintzmann, Rainer},
	urldate = {2021-07-15},
	date = {2021-06-14},
	note = {Publisher: Royal Society},
}

@article{shah_deep-learning_2021,
	title = {Deep-learning based denoising and reconstruction of super-resolution structured illumination microscopy images},
	issn = {2327-9125},
	url = {https://www.osapublishing.org/prj/abstract.cfm?msid=},
	doi = {10.1364/PRJ.416437},
	journaltitle = {Photonics Research},
	author = {Shah, Zafran Hussain and Müller, Marcel and Wang, Tung-Cheng and Scheidig, Philip and Schneider, Axel and Schüttpelz, Mark and Huser, Thomas and Schenck, Wolfram},
	urldate = {2021-04-12},
	date = {2021-01-29},
	note = {Publisher: Optical Society of America},
}

@article{qiao_evaluation_2021,
	title = {Evaluation and development of deep neural networks for image super-resolution in optical microscopy},
	volume = {18},
	rights = {2021 The Author(s), under exclusive licence to Springer Nature America, Inc.},
	issn = {1548-7105},
	url = {https://www.nature.com/articles/s41592-020-01048-5},
	doi = {10.1038/s41592-020-01048-5},
	pages = {194--202},
	number = {2},
	journaltitle = {Nature Methods},
	author = {Qiao, Chang and Li, Di and Guo, Yuting and Liu, Chong and Jiang, Tao and Dai, Qionghai and Li, Dong},
	urldate = {2021-04-12},
	date = {2021-02},
	langid = {english},
	note = {Number: 2
Publisher: Nature Publishing Group},
}

@article{jin_deep_2020,
	title = {Deep learning enables structured illumination microscopy with low light levels and enhanced speed},
	volume = {11},
	rights = {2020 The Author(s)},
	issn = {2041-1723},
	url = {https://www.nature.com/articles/s41467-020-15784-x},
	doi = {10.1038/s41467-020-15784-x},
	pages = {1934},
	number = {1},
	journaltitle = {Nature Communications},
	author = {Jin, Luhong and Liu, Bei and Zhao, Fenqiang and Hahn, Stephen and Dong, Bowei and Song, Ruiyan and Elston, Timothy C. and Xu, Yingke and Hahn, Klaus M.},
	urldate = {2021-04-12},
	date = {2020-04-22},
	langid = {english},
	note = {Number: 1
Publisher: Nature Publishing Group},
}

@article{ling_fast_2020,
	title = {Fast structured illumination microscopy via deep learning},
	volume = {8},
	rights = {\&\#169; 2020 Chinese Laser Press},
	issn = {2327-9125},
	url = {https://www.osapublishing.org/prj/abstract.cfm?uri=prj-8-8-1350},
	doi = {10.1364/PRJ.396122},
	pages = {1350--1359},
	number = {8},
	journaltitle = {Photonics Research},
	shortjournal = {Photon. Res., {PRJ}},
	author = {Ling, Chang and Zhang, Chonglei and Zhang, Chonglei and Wang, Mingqun and Meng, Fanfei and Du, Luping and Du, Luping and Yuan, Xiaocong and Yuan, Xiaocong},
	urldate = {2021-04-12},
	date = {2020-08-01},
	note = {Publisher: Optical Society of America},
}

@article{prakash_super-resolution_2021,
	title = {Super-resolution structured illumination microscopy: past, present and future},
	url = {http://arxiv.org/abs/2102.13649},
	shorttitle = {Super-resolution structured illumination microscopy},
	journaltitle = {{arXiv}:2102.13649 [physics, q-bio]},
	author = {Prakash, Kirti and Diederich, Benedict and Reichelt, Stefanie and Heintzmann, Rainer and Schermelleh, Lothar},
	urldate = {2021-04-12},
	date = {2021-02-26},
	eprinttype = {arxiv},
	eprint = {2102.13649},
}

@article{gustafsson_surpassing_2000,
	title = {Surpassing the lateral resolution limit by a factor of two using structured illumination microscopy},
	volume = {198},
	issn = {1365-2818},
	url = {http://onlinelibrary.wiley.com/doi/10.1046/j.1365-2818.2000.00710.x/abstract},
	doi = {10.1046/j.1365-2818.2000.00710.x},
	pages = {82--87},
	number = {2},
	journaltitle = {Journal of Microscopy},
	author = {Gustafsson, Mats G. L.},
	urldate = {2015-12-14},
	date = {2000-05-01},
}

@article{neil_method_1997,
	title = {Method of obtaining optical sectioning by using structured light in a conventional microscope},
	volume = {22},
	rights = {© 1997 Optical Society of America},
	issn = {1539-4794},
	url = {https://www.osapublishing.org/abstract.cfm?uri=ol-22-24-1905},
	doi = {10.1364/OL.22.001905},
	pages = {1905--1907},
	number = {24},
	journaltitle = {Optics Letters},
	shortjournal = {Opt. Lett., {OL}},
	author = {Neil, Mark and Juškaitis, R. and Wilson, T.},
	urldate = {2017-09-12},
	date = {1997-12-15},
}

@article{prabhat_simultaneous_2004,
	title = {Simultaneous imaging of different focal planes in fluorescence microscopy for the study of cellular dynamics in three dimensions},
	volume = {3},
	issn = {1558-2639},
	doi = {10.1109/TNB.2004.837899},
	pages = {237--242},
	number = {4},
	journaltitle = {{IEEE} Transactions on {NanoBioscience}},
	author = {Prabhat, P. and Ram, S. and Ward, E. S. and Ober, R. J.},
	date = {2004-12},
	note = {Conference Name: {IEEE} Transactions on {NanoBioscience}},
}

@article{descloux_high-speed_2020,
	title = {High-speed multiplane structured illumination microscopy of living cells using an image-splitting prism},
	volume = {9},
	issn = {2192-8614},
	url = {https://www.degruyter.com/document/doi/10.1515/nanoph-2019-0346/html},
	doi = {10.1515/nanoph-2019-0346},
	pages = {143--148},
	number = {1},
	journaltitle = {Nanophotonics},
	author = {Descloux, Adrien and Müller, Marcel and Navikas, Vytautas and Markwirth, Andreas and Eynde, Robin van den and Lukes, Tomas and Hübner, Wolfgang and Lasser, Theo and Radenovic, Aleksandra and Dedecker, Peter and Huser, Thomas},
	urldate = {2021-04-12},
	date = {2020-01-01},
	langid = {english},
	note = {Publisher: De Gruyter
Section: Nanophotonics},
}

@article{ball_simcheck_2015,
	title = {{SIMcheck}: a Toolbox for Successful Super-resolution Structured Illumination Microscopy},
	volume = {5},
	rights = {2015 The Author(s)},
	issn = {2045-2322},
	url = {https://www.nature.com/articles/srep15915},
	doi = {10.1038/srep15915},
	shorttitle = {{SIMcheck}},
	pages = {15915},
	number = {1},
	journaltitle = {Scientific Reports},
	author = {Ball, Graeme and Demmerle, Justin and Kaufmann, Rainer and Davis, Ilan and Dobbie, Ian M. and Schermelleh, Lothar},
	urldate = {2021-04-12},
	date = {2015-11-03},
	langid = {english},
	note = {Number: 1
Publisher: Nature Publishing Group},
}

@article{barbieri_two-dimensional_2021,
	title = {Two-dimensional {TIRF}-{SIM}–traction force microscopy (2D {TIRF}-{SIM}-{TFM})},
	volume = {12},
	rights = {2021 The Author(s)},
	issn = {2041-1723},
	url = {https://www.nature.com/articles/s41467-021-22377-9},
	doi = {10.1038/s41467-021-22377-9},
	pages = {2169},
	number = {1},
	journaltitle = {Nature Communications},
	author = {Barbieri, Liliana and Colin-York, Huw and Korobchevskaya, Kseniya and Li, Di and Wolfson, Deanna L. and Karedla, Narain and Schneider, Falk and Ahluwalia, Balpreet S. and Seternes, Tore and Dalmo, Roy A. and Dustin, Michael L. and Li, Dong and Fritzsche, Marco},
	urldate = {2021-04-12},
	date = {2021-04-12},
	langid = {english},
	note = {Number: 1
Publisher: Nature Publishing Group},
}

@article{colin-york_spatiotemporally_2019,
	title = {Spatiotemporally Super-Resolved Volumetric Traction Force Microscopy},
	volume = {19},
	issn = {1530-6984},
	url = {https://doi.org/10.1021/acs.nanolett.9b01196},
	doi = {10.1021/acs.nanolett.9b01196},
	pages = {4427--4434},
	number = {7},
	journaltitle = {Nano Letters},
	shortjournal = {Nano Lett.},
	author = {Colin-York, Huw and Javanmardi, Yousef and Barbieri, Liliana and Li, Di and Korobchevskaya, Kseniya and Guo, Yuting and Hall, Chloe and Taylor, Aaron and Khuon, Satya and Sheridan, Graham K. and Chew, Teng-Leong and Li, Dong and Moeendarbary, Emad and Fritzsche, Marco},
	urldate = {2021-04-12},
	date = {2019-07-10},
	note = {Publisher: American Chemical Society},
}

@article{gorlitz_mapping_2017,
	title = {Mapping Molecular Function to Biological Nanostructure: Combining Structured Illumination Microscopy with Fluorescence Lifetime Imaging ({SIM} + {FLIM})},
	volume = {4},
	rights = {http://creativecommons.org/licenses/by/3.0/},
	url = {https://www.mdpi.com/2304-6732/4/3/40},
	doi = {10.3390/photonics4030040},
	shorttitle = {Mapping Molecular Function to Biological Nanostructure},
	pages = {40},
	number = {3},
	journaltitle = {Photonics},
	author = {Görlitz, Frederik and Corcoran, David S. and Garcia Castano, Edwin A. and Leitinger, Birgit and Neil, Mark A. A. and Dunsby, Christopher and French, Paul M. W.},
	urldate = {2021-04-12},
	date = {2017-09},
	langid = {english},
	note = {Number: 3
Publisher: Multidisciplinary Digital Publishing Institute},
}

@article{hinsdale_optically_2017,
	title = {Optically sectioned wide-field fluorescence lifetime imaging microscopy enabled by structured illumination},
	volume = {8},
	rights = {\&\#169; 2017 Optical Society of America},
	issn = {2156-7085},
	url = {https://www.osapublishing.org/boe/abstract.cfm?uri=boe-8-3-1455},
	doi = {10.1364/BOE.8.001455},
	pages = {1455--1465},
	number = {3},
	journaltitle = {Biomedical Optics Express},
	shortjournal = {Biomed. Opt. Express, {BOE}},
	author = {Hinsdale, Taylor and Olsovsky, Cory and Rico-Jimenez, Jose J. and Maitland, Kristen C. and Jo, Javier A. and Malik, Bilal H.},
	urldate = {2021-04-12},
	date = {2017-03-01},
	note = {Publisher: Optical Society of America},
}

@article{zhanghao_super-resolution_2019,
	title = {Super-resolution imaging of fluorescent dipoles via polarized structured illumination microscopy},
	volume = {10},
	rights = {2019 The Author(s)},
	issn = {2041-1723},
	url = {https://www.nature.com/articles/s41467-019-12681-w},
	doi = {10.1038/s41467-019-12681-w},
	pages = {4694},
	number = {1},
	journaltitle = {Nature Communications},
	author = {Zhanghao, Karl and Chen, Xingye and Liu, Wenhui and Li, Meiqi and Liu, Yiqiong and Wang, Yiming and Luo, Sha and Wang, Xiao and Shan, Chunyan and Xie, Hao and Gao, Juntao and Chen, Xiaowei and Jin, Dayong and Li, Xiangdong and Zhang, Yan and Dai, Qionghai and Xi, Peng},
	urldate = {2021-04-12},
	date = {2019-10-16},
	langid = {english},
	note = {Number: 1
Publisher: Nature Publishing Group},
}

@article{yeh_structured_2017,
	title = {Structured illumination microscopy with unknown patterns and a statistical prior},
	volume = {8},
	rights = {\&\#169; 2017 Optical Society of America},
	issn = {2156-7085},
	url = {https://www.osapublishing.org/boe/abstract.cfm?uri=boe-8-2-695},
	doi = {10.1364/BOE.8.000695},
	pages = {695--711},
	number = {2},
	journaltitle = {Biomedical Optics Express},
	shortjournal = {Biomed. Opt. Express, {BOE}},
	author = {Yeh, Li-Hao and Tian, Lei and Waller, Laura},
	urldate = {2021-04-12},
	date = {2017-02-01},
	note = {Publisher: Optical Society of America},
}

@article{ayuk_structured_2013,
	title = {Structured illumination fluorescence microscopy with distorted excitations using a filtered blind-{SIM} algorithm},
	volume = {38},
	rights = {\&\#169; 2013 Optical Society of America},
	issn = {1539-4794},
	url = {https://www.osapublishing.org/ol/abstract.cfm?uri=ol-38-22-4723},
	doi = {10.1364/OL.38.004723},
	pages = {4723--4726},
	number = {22},
	journaltitle = {Optics Letters},
	shortjournal = {Opt. Lett., {OL}},
	author = {Ayuk, R. and Giovannini, H. and Jost, A. and Mudry, E. and Girard, J. and Mangeat, T. and Sandeau, N. and Heintzmann, R. and Wicker, K. and Belkebir, K. and Sentenac, A.},
	urldate = {2021-04-12},
	date = {2013-11-15},
	note = {Publisher: Optical Society of America},
}

@article{jost_optical_2015,
	title = {Optical Sectioning and High Resolution in Single-Slice Structured Illumination Microscopy by Thick Slice Blind-{SIM} Reconstruction},
	volume = {10},
	issn = {1932-6203},
	url = {https://journals.plos.org/plosone/article?id=10.1371/journal.pone.0132174},
	doi = {10.1371/journal.pone.0132174},
	pages = {e0132174},
	number = {7},
	journaltitle = {{PLOS} {ONE}},
	shortjournal = {{PLOS} {ONE}},
	author = {Jost, Aurélie and Tolstik, Elen and Feldmann, Polina and Wicker, Kai and Sentenac, Anne and Heintzmann, Rainer},
	urldate = {2021-04-12},
	date = {2015-07-06},
	langid = {english},
	note = {Publisher: Public Library of Science},
}

@article{huang_fast_2018,
	title = {Fast, long-term, super-resolution imaging with Hessian structured illumination microscopy},
	volume = {36},
	rights = {2018 Nature Publishing Group, a division of Macmillan Publishers Limited. All Rights Reserved.},
	issn = {1546-1696},
	url = {https://www.nature.com/articles/nbt.4115},
	doi = {10.1038/nbt.4115},
	pages = {451--459},
	number = {5},
	journaltitle = {Nature Biotechnology},
	author = {Huang, Xiaoshuai and Fan, Junchao and Li, Liuju and Liu, Haosen and Wu, Runlong and Wu, Yi and Wei, Lisi and Mao, Heng and Lal, Amit and Xi, Peng and Tang, Liqiang and Zhang, Yunfeng and Liu, Yanmei and Tan, Shan and Chen, Liangyi},
	urldate = {2021-04-12},
	date = {2018-05},
	langid = {english},
	note = {Number: 5
Publisher: Nature Publishing Group},
}

@article{boulanger_nonsmooth_2018,
	title = {Nonsmooth convex optimization for structured illumination microscopy image reconstruction},
	volume = {34},
	issn = {0266-5611},
	url = {https://doi.org/10.1088/1361-6420/aaccca},
	doi = {10.1088/1361-6420/aaccca},
	pages = {095004},
	number = {9},
	journaltitle = {Inverse Problems},
	shortjournal = {Inverse Problems},
	author = {Boulanger, Jérôme and Pustelnik, Nelly and Condat, Laurent and Sengmanivong, Lucie and Piolot, Tristan},
	urldate = {2021-04-12},
	date = {2018-07},
	langid = {english},
	note = {Publisher: {IOP} Publishing},
}

@article{huff_airyscan_2015,
	title = {The Airyscan detector from {ZEISS}: confocal imaging with improved signal-to-noise ratio and super-resolution},
	volume = {12},
	rights = {2015 Nature Publishing Group, a division of Macmillan Publishers Limited. All Rights Reserved.},
	issn = {1548-7105},
	url = {https://www.nature.com/articles/nmeth.f.388},
	doi = {10.1038/nmeth.f.388},
	shorttitle = {The Airyscan detector from {ZEISS}},
	pages = {i--ii},
	number = {12},
	journaltitle = {Nature Methods},
	author = {Huff, Joseph},
	urldate = {2021-04-12},
	date = {2015-12},
	langid = {english},
	note = {Number: 12
Publisher: Nature Publishing Group},
}

@article{schmidt_cell_2018,
	title = {Cell Detection with Star-convex Polygons},
	volume = {11071},
	url = {http://arxiv.org/abs/1806.03535},
	doi = {10.1007/978-3-030-00934-2_30},
	pages = {265--273},
	journaltitle = {{arXiv}},
	author = {Schmidt, Uwe and Weigert, Martin and Broaddus, Coleman and Myers, Gene},
	urldate = {2021-04-06},
	date = {2018},
	eprinttype = {arxiv},
	eprint = {1806.03535},
}

@article{winter_two-photon_2014,
	title = {Two-photon instant structured illumination microscopy improves the depth penetration of super-resolution imaging in thick scattering samples},
	volume = {1},
	issn = {2334-2536},
	url = {https://www.osapublishing.org/optica/abstract.cfm?uri=optica-1-3-181},
	doi = {10.1364/OPTICA.1.000181},
	pages = {181--191},
	number = {3},
	journaltitle = {Optica},
	shortjournal = {Optica, {OPTICA}},
	author = {Winter, Peter W. and York, Andrew G. and Nogare, Damian Dalle and Ingaramo, Maria and Christensen, Ryan and Chitnis, Ajay and Patterson, George H. and Shroff, Hari},
	urldate = {2021-04-12},
	date = {2014-09-20},
	note = {Publisher: Optical Society of America},
}

@article{demmerle_strategic_2017,
	title = {Strategic and practical guidelines for successful structured illumination microscopy},
	volume = {12},
	rights = {2017 Nature Publishing Group, a division of Macmillan Publishers Limited. All Rights Reserved.},
	issn = {1750-2799},
	url = {https://www.nature.com/articles/nprot.2017.019},
	doi = {10.1038/nprot.2017.019},
	pages = {988--1010},
	number = {5},
	journaltitle = {Nature Protocols},
	author = {Demmerle, Justin and Innocent, Cassandravictoria and North, Alison J. and Ball, Graeme and Müller, Marcel and Miron, Ezequiel and Matsuda, Atsushi and Dobbie, Ian M. and Markaki, Yolanda and Schermelleh, Lothar},
	urldate = {2021-04-08},
	date = {2017-05},
	langid = {english},
	note = {Number: 5
Publisher: Nature Publishing Group},
}

@article{azuma_super-resolution_2015,
	title = {Super-resolution spinning-disk confocal microscopy using optical photon reassignment},
	volume = {23},
	rights = {\&\#169; 2015 Optical Society of America},
	issn = {1094-4087},
	url = {https://www.osapublishing.org/oe/abstract.cfm?uri=oe-23-11-15003},
	doi = {10.1364/OE.23.015003},
	pages = {15003--15011},
	number = {11},
	journaltitle = {Optics Express},
	shortjournal = {Opt. Express, {OE}},
	author = {Azuma, Takuya and Kei, Takayuki},
	urldate = {2021-04-08},
	date = {2015-06-01},
	note = {Publisher: Optical Society of America},
}

@article{ingerman_signal_2019,
	title = {Signal, noise and resolution in linear and nonlinear structured-illumination microscopy},
	volume = {273},
	issn = {1365-2818},
	url = {https://onlinelibrary.wiley.com/doi/abs/10.1111/jmi.12753},
	doi = {https://doi.org/10.1111/jmi.12753},
	pages = {3--25},
	number = {1},
	journaltitle = {Journal of Microscopy},
	author = {Ingerman, E. A. and London, R. A. and Heintzmann, R. and Gustafsson, M. G. L.},
	urldate = {2021-04-07},
	date = {2019},
	langid = {english},
	note = {\_eprint: https://onlinelibrary.wiley.com/doi/pdf/10.1111/jmi.12753},
}

@article{dasgupta_direct_2021,
	title = {Direct supercritical angle localization microscopy for nanometer 3D superresolution},
	volume = {12},
	rights = {2021 The Author(s)},
	issn = {2041-1723},
	url = {https://www.nature.com/articles/s41467-021-21333-x},
	doi = {10.1038/s41467-021-21333-x},
	pages = {1180},
	number = {1},
	journaltitle = {Nature Communications},
	author = {Dasgupta, Anindita and Deschamps, Joran and Matti, Ulf and Hübner, Uwe and Becker, Jan and Strauss, Sebastian and Jungmann, Ralf and Heintzmann, Rainer and Ries, Jonas},
	urldate = {2021-04-07},
	date = {2021-02-19},
	langid = {english},
	note = {Number: 1
Publisher: Nature Publishing Group},
}

@article{liu_cell_2016,
	title = {Cell refractive index for cell biology and disease diagnosis: past, present and future},
	volume = {16},
	issn = {1473-0189},
	url = {https://pubs.rsc.org/en/content/articlelanding/2016/lc/c5lc01445j},
	doi = {10.1039/C5LC01445J},
	shorttitle = {Cell refractive index for cell biology and disease diagnosis},
	pages = {634--644},
	number = {4},
	journaltitle = {Lab on a Chip},
	shortjournal = {Lab Chip},
	author = {Liu, P. Y. and Chin, L. K. and Ser, W. and Chen, H. F. and Hsieh, C.-M. and Lee, C.-H. and Sung, K.-B. and Ayi, T. C. and Yap, P. H. and Liedberg, B. and Wang, K. and Bourouina, T. and Leprince-Wang, Y.},
	urldate = {2021-04-07},
	date = {2016-02-09},
	langid = {english},
	note = {Publisher: The Royal Society of Chemistry},
}

@article{ewald_zur_1913,
	title = {Zur Theorie der Interferenzen der Röntgentstrahlen in Kristallen},
	volume = {11},
	pages = {465--472},
	journaltitle = {Physikalische Zeitschrift},
	author = {Ewald, P.P.},
	date = {1913-06-01},
}

@article{mccutchen_generalized_1964,
	title = {Generalized Aperture and the Three-Dimensional Diffraction Image},
	volume = {54},
	rights = {\&\#169; 1964 Optical Society of America},
	url = {https://www.osapublishing.org/josa/abstract.cfm?uri=josa-54-2-240},
	doi = {10.1364/JOSA.54.000240},
	pages = {240--244},
	number = {2},
	journaltitle = {{JOSA}},
	shortjournal = {J. Opt. Soc. Am., {JOSA}},
	author = {{McCutchen}, C. W.},
	urldate = {2021-04-07},
	date = {1964-02-01},
	note = {Publisher: Optical Society of America},
}

@article{smith_structured_2021,
	title = {Structured illumination microscopy with noise-controlled image reconstructions},
	rights = {© 2021, Posted by Cold Spring Harbor Laboratory. This pre-print is available under a Creative Commons License (Attribution-{NoDerivs} 4.0 International), {CC} {BY}-{ND} 4.0, as described at http://creativecommons.org/licenses/by-nd/4.0/},
	url = {https://www.biorxiv.org/content/10.1101/2021.03.11.434940v1},
	doi = {10.1101/2021.03.11.434940},
	pages = {2021.03.11.434940},
	journaltitle = {{bioRxiv}},
	author = {Smith, Carlas S. and Slotman, Johan A. and Schermelleh, Lothar and Chakrova, Nadya and Hari, Sangeetha and Vos, Yoram and Hagen, Cornelis W. and Müller, Marcel and Cappellen, Wiggert van and Houtsmuller, Adriaan B. and Hoogenboom, Jacob P. and Stallinga, Sjoerd},
	urldate = {2021-04-06},
	date = {2021-03-11},
	langid = {english},
	note = {Publisher: Cold Spring Harbor Laboratory
Section: New Results},
}

@article{lahrberg_accurate_2018,
	title = {Accurate estimation of the illumination pattern’s orientation and wavelength in sinusoidal structured illumination microscopy},
	volume = {57},
	rights = {\&\#169; 2018 Optical Society of America},
	issn = {2155-3165},
	url = {https://www.osapublishing.org/ao/abstract.cfm?uri=ao-57-5-1019},
	doi = {10.1364/AO.57.001019},
	pages = {1019--1025},
	number = {5},
	journaltitle = {Applied Optics},
	shortjournal = {Appl. Opt., {AO}},
	author = {Lahrberg, Marcel and Singh, Mandeep and Khare, Kedar and Ahluwalia, Balpreet Singh},
	urldate = {2021-04-06},
	date = {2018-02-10},
	note = {Publisher: Optical Society of America},
}

@article{guo_rapid_2020,
	title = {Rapid image deconvolution and multiview fusion for optical microscopy},
	volume = {38},
	rights = {2020 The Author(s), under exclusive licence to Springer Nature America, Inc.},
	issn = {1546-1696},
	url = {https://www.nature.com/articles/s41587-020-0560-x},
	doi = {10.1038/s41587-020-0560-x},
	pages = {1337--1346},
	number = {11},
	journaltitle = {Nature Biotechnology},
	author = {Guo, Min and Li, Yue and Su, Yijun and Lambert, Talley and Nogare, Damian Dalle and Moyle, Mark W. and Duncan, Leighton H. and Ikegami, Richard and Santella, Anthony and Rey-Suarez, Ivan and Green, Daniel and Beiriger, Anastasia and Chen, Jiji and Vishwasrao, Harshad and Ganesan, Sundar and Prince, Victoria and Waters, Jennifer C. and Annunziata, Christina M. and Hafner, Markus and Mohler, William A. and Chitnis, Ajay B. and Upadhyaya, Arpita and Usdin, Ted B. and Bao, Zhirong and Colón-Ramos, Daniel and La Riviere, Patrick and Liu, Huafeng and Wu, Yicong and Shroff, Hari},
	urldate = {2021-04-06},
	date = {2020-11},
	langid = {english},
	note = {Number: 11
Publisher: Nature Publishing Group},
}

@article{arganda-carreras_trainable_2017,
	title = {Trainable Weka Segmentation: a machine learning tool for microscopy pixel classification},
	volume = {33},
	issn = {1367-4803},
	url = {https://doi.org/10.1093/bioinformatics/btx180},
	doi = {10.1093/bioinformatics/btx180},
	shorttitle = {Trainable Weka Segmentation},
	pages = {2424--2426},
	number = {15},
	journaltitle = {Bioinformatics},
	shortjournal = {Bioinformatics},
	author = {Arganda-Carreras, Ignacio and Kaynig, Verena and Rueden, Curtis and Eliceiri, Kevin W and Schindelin, Johannes and Cardona, Albert and Sebastian Seung, H},
	urldate = {2021-04-06},
	date = {2017-08-01},
}

@article{boulanger_patch-based_2010,
	title = {Patch-Based Nonlocal Functional for Denoising Fluorescence Microscopy Image Sequences},
	volume = {29},
	issn = {1558-254X},
	doi = {10.1109/TMI.2009.2033991},
	pages = {442--454},
	number = {2},
	journaltitle = {{IEEE} Transactions on Medical Imaging},
	author = {Boulanger, J. and Kervrann, C. and Bouthemy, P. and Elbau, P. and Sibarita, J. and Salamero, J.},
	date = {2010-02},
	note = {Conference Name: {IEEE} Transactions on Medical Imaging},
}

@article{orieux_bayesian_2012,
	title = {Bayesian Estimation for Optimized Structured Illumination Microscopy},
	volume = {21},
	issn = {1941-0042},
	doi = {10.1109/TIP.2011.2162741},
	pages = {601--614},
	number = {2},
	journaltitle = {{IEEE} Transactions on Image Processing},
	author = {Orieux, F. and Sepulveda, E. and Loriette, V. and Dubertret, B. and Olivo-Marin, J.},
	date = {2012-02},
	note = {Conference Name: {IEEE} Transactions on Image Processing},
}

@article{lal_frequency_2018,
	title = {A Frequency Domain {SIM} Reconstruction Algorithm Using Reduced Number of Images},
	volume = {27},
	issn = {1941-0042},
	doi = {10.1109/TIP.2018.2842149},
	pages = {4555--4570},
	number = {9},
	journaltitle = {{IEEE} Transactions on Image Processing},
	author = {Lal, A. and Shan, C. and Zhao, K. and Liu, W. and Huang, X. and Zong, W. and Chen, L. and Xi, P.},
	date = {2018-09},
	note = {Conference Name: {IEEE} Transactions on Image Processing},
}

@article{ma_structured_2018,
	title = {Structured illumination microscopy with interleaved reconstruction ({SIMILR})},
	volume = {11},
	rights = {© 2017 {WILEY}‐{VCH} Verlag {GmbH} \& Co. {KGaA}, Weinheim},
	issn = {1864-0648},
	url = {https://onlinelibrary.wiley.com/doi/abs/10.1002/jbio.201700090},
	doi = {https://doi.org/10.1002/jbio.201700090},
	pages = {e201700090},
	number = {2},
	journaltitle = {Journal of Biophotonics},
	author = {Ma, Ying and Li, Di and Smith, Zachary J. and Li, Dong and Chu, Kaiqin},
	urldate = {2021-04-06},
	date = {2018},
	langid = {english},
	note = {\_eprint: https://onlinelibrary.wiley.com/doi/pdf/10.1002/jbio.201700090},
}

@article{boualam_method_2021,
	title = {Method for assessing the spatiotemporal resolution of structured illumination microscopy ({SIM})},
	volume = {12},
	issn = {2156-7085},
	url = {https://www.osapublishing.org/boe/abstract.cfm?uri=boe-12-2-790},
	doi = {10.1364/BOE.403592},
	pages = {790--801},
	number = {2},
	journaltitle = {Biomedical Optics Express},
	shortjournal = {Biomed. Opt. Express, {BOE}},
	author = {Boualam, Abderrahim and Rowlands, Christopher J.},
	urldate = {2021-04-06},
	date = {2021-02-01},
	note = {Publisher: Optical Society of America},
}

@article{markwirth_video-rate_2019,
	title = {Video-rate multi-color structured illumination microscopy with simultaneous real-time reconstruction},
	volume = {10},
	rights = {2019 The Author(s)},
	issn = {2041-1723},
	url = {https://www.nature.com/articles/s41467-019-12165-x},
	doi = {10.1038/s41467-019-12165-x},
	pages = {4315},
	number = {1},
	journaltitle = {Nature Communications},
	author = {Markwirth, Andreas and Lachetta, Mario and Mönkemöller, Viola and Heintzmann, Rainer and Hübner, Wolfgang and Huser, Thomas and Müller, Marcel},
	urldate = {2021-04-06},
	date = {2019-09-20},
	langid = {english},
	note = {Number: 1
Publisher: Nature Publishing Group},
}

@article{guo_visualizing_2018,
	title = {Visualizing Intracellular Organelle and Cytoskeletal Interactions at Nanoscale Resolution on Millisecond Timescales},
	volume = {175},
	issn = {0092-8674},
	url = {https://www.sciencedirect.com/science/article/pii/S0092867418313084},
	doi = {10.1016/j.cell.2018.09.057},
	pages = {1430--1442.e17},
	number = {5},
	journaltitle = {Cell},
	shortjournal = {Cell},
	author = {Guo, Yuting and Li, Di and Zhang, Siwei and Yang, Yanrui and Liu, Jia-Jia and Wang, Xinyu and Liu, Chong and Milkie, Daniel E. and Moore, Regan P. and Tulu, U. Serdar and Kiehart, Daniel P. and Hu, Junjie and Lippincott-Schwartz, Jennifer and Betzig, Eric and Li, Dong},
	urldate = {2021-04-06},
	date = {2018-11-15},
	langid = {english},
}

@article{pospisil_highly_2021,
	title = {Highly compact and cost-effective 2-beam super-resolution structured illumination microscope based on all-fiber optic components},
	volume = {29},
	rights = {\&\#169; 2021 Optical Society of America},
	issn = {1094-4087},
	url = {https://www.osapublishing.org/oe/abstract.cfm?uri=oe-29-8-11833},
	doi = {10.1364/OE.420592},
	pages = {11833--11844},
	number = {8},
	journaltitle = {Optics Express},
	shortjournal = {Opt. Express, {OE}},
	author = {Pospíšil, Jakub and Pospíšil, Jakub and Wiebusch, Gerd and Fliegel, Karel and Klíma, Miloš and Huser, Thomas},
	urldate = {2021-04-06},
	date = {2021-04-12},
	note = {Publisher: Optical Society of America},
}

@article{hinsdale_high-speed_2021,
	title = {High-speed multicolor structured illumination microscopy using a hexagonal single mode fiber array},
	volume = {12},
	rights = {\&\#169; 2021 Optical Society of America},
	issn = {2156-7085},
	url = {https://www.osapublishing.org/boe/abstract.cfm?uri=boe-12-2-1181},
	doi = {10.1364/BOE.416546},
	pages = {1181--1194},
	number = {2},
	journaltitle = {Biomedical Optics Express},
	shortjournal = {Biomed. Opt. Express, {BOE}},
	author = {Hinsdale, Taylor A. and Stallinga, Sjoerd and Rieger, Bernd},
	urldate = {2021-04-06},
	date = {2021-02-01},
	note = {Publisher: Optical Society of America},
}

@article{brown_multicolor_2020,
	title = {Multicolor structured illumination microscopy and quantitative control of polychromatic coherent light with a digital micromirror device},
	rights = {© 2020, Posted by Cold Spring Harbor Laboratory. This pre-print is available under a Creative Commons License (Attribution-{NonCommercial} 4.0 International), {CC} {BY}-{NC} 4.0, as described at http://creativecommons.org/licenses/by-nc/4.0/},
	url = {https://www.biorxiv.org/content/10.1101/2020.07.27.223941v3},
	doi = {10.1101/2020.07.27.223941},
	pages = {2020.07.27.223941},
	journaltitle = {{bioRxiv}},
	author = {Brown, Peter T. and Kruithoff, Rory and Seedorf, Gregory J. and Shepherd, Douglas P.},
	urldate = {2021-04-06},
	date = {2020-12-08},
	langid = {english},
	note = {Publisher: Cold Spring Harbor Laboratory
Section: New Results},
}

@article{sandmeyer_dmd-based_2019,
	title = {{DMD}-based super-resolution structured illumination microscopy visualizes live cell dynamics at high speed and low cost},
	rights = {© 2019, Posted by Cold Spring Harbor Laboratory. The copyright holder for this pre-print is the author. All rights reserved. The material may not be redistributed, re-used or adapted without the author's permission.},
	url = {https://www.biorxiv.org/content/10.1101/797670v1},
	doi = {10.1101/797670},
	pages = {797670},
	journaltitle = {{bioRxiv}},
	author = {Sandmeyer, Alice and Lachetta, Mario and Sandmeyer, Hauke and Hübner, Wolfgang and Huser, Thomas and Müller, Marcel},
	urldate = {2021-04-06},
	date = {2019-10-08},
	langid = {english},
	note = {Publisher: Cold Spring Harbor Laboratory
Section: New Results},
}

@patent{dougherty_method_2013,
	title = {Method and system for fast three-dimensional structured-illumination-microscopy imaging},
	url = {https://patents.google.com/patent/US8570650B2/en?q=METHOD+AND+SYSTEM+FOR+FAST+THREE-DIMENSIONAL+STRUCTURED-ILLUMINATION-MICROSCOPY+IMAGING&oq=METHOD+AND+SYSTEM+FOR+FAST+THREE-DIMENSIONAL+STRUCTURED-ILLUMINATION-MICROSCOPY+IMAGING},
	holder = {Applied Precision Inc},
	type = {patentus},
	number = {8570650B2},
	author = {Dougherty, William M. and Quarre, Steven Charles},
	urldate = {2021-04-06},
	date = {2013-10-29},
	langid = {english},
}

@article{best_structured_2011,
	title = {Structured illumination microscopy of autofluorescent aggregations in human tissue},
	volume = {42},
	issn = {0968-4328},
	url = {https://www.sciencedirect.com/science/article/pii/S096843281000212X},
	doi = {10.1016/j.micron.2010.06.016},
	series = {Special issue on Super and High resolution imaging},
	pages = {330--335},
	number = {4},
	journaltitle = {Micron},
	shortjournal = {Micron},
	author = {Best, Gerrit and Amberger, Roman and Baddeley, David and Ach, Thomas and Dithmar, Stefan and Heintzmann, Rainer and Cremer, Christoph},
	urldate = {2021-04-06},
	date = {2011-06-01},
	langid = {english},
}

@article{liu_three-dimensional_2019,
	title = {Three-dimensional super-resolution imaging of live whole cells using galvanometer-based structured illumination microscopy},
	volume = {27},
	rights = {\&\#169; 2019 Optical Society of America},
	issn = {1094-4087},
	url = {https://www.osapublishing.org/oe/abstract.cfm?uri=oe-27-5-7237},
	doi = {10.1364/OE.27.007237},
	pages = {7237--7248},
	number = {5},
	journaltitle = {Optics Express},
	shortjournal = {Opt. Express, {OE}},
	author = {Liu, Wenjie and Liu, Qiulan and Zhang, Zhimin and Han, Yubing and Kuang, Cuifang and Xu, Liang and Yang, Hongqin and Liu, Xu},
	urldate = {2021-04-06},
	date = {2019-03-04},
	note = {Publisher: Optical Society of America},
}

@article{manton_concepts_2020,
	title = {Concepts for structured illumination microscopy with extended axial resolution through mirrored illumination},
	volume = {11},
	issn = {2156-7085},
	url = {https://www.osapublishing.org/boe/abstract.cfm?uri=boe-11-4-2098},
	doi = {10.1364/BOE.382398},
	pages = {2098--2108},
	number = {4},
	journaltitle = {Biomedical Optics Express},
	shortjournal = {Biomed. Opt. Express, {BOE}},
	author = {Manton, James D. and Manton, James D. and Ströhl, Florian and Ströhl, Florian and Fiolka, Reto and Fiolka, Reto and Kaminski, Clemens F. and Rees, Eric J.},
	urldate = {2021-04-06},
	date = {2020-04-01},
	note = {Publisher: Optical Society of America},
}

@article{reymond_simple_2019,
	title = {{SIMPLE}: Structured illumination based point localization estimator with enhanced precision},
	volume = {27},
	rights = {\&\#169; 2019 Optical Society of America},
	issn = {1094-4087},
	url = {https://www.osapublishing.org/oe/abstract.cfm?uri=oe-27-17-24578},
	doi = {10.1364/OE.27.024578},
	shorttitle = {{SIMPLE}},
	pages = {24578--24590},
	number = {17},
	journaltitle = {Optics Express},
	shortjournal = {Opt. Express, {OE}},
	author = {Reymond, Loïc and Reymond, Loïc and Ziegler, Johannes and Knapp, Christian and Wang, Fung-Chen and Huser, Thomas and Ruprecht, Verena and Ruprecht, Verena and Wieser, Stefan},
	urldate = {2021-04-06},
	date = {2019-08-19},
	note = {Publisher: Optical Society of America},
}

@article{schmidt_camera-based_2021,
	title = {Camera-based localization microscopy optimized with calibrated structured illumination},
	volume = {4},
	rights = {2021 The Author(s)},
	issn = {2399-3650},
	url = {https://www.nature.com/articles/s42005-021-00546-y},
	doi = {10.1038/s42005-021-00546-y},
	pages = {1--9},
	number = {1},
	journaltitle = {Communications Physics},
	author = {Schmidt, Martin and Hundahl, Adam C. and Flyvbjerg, Henrik and Marie, Rodolphe and Mortensen, Kim I.},
	urldate = {2021-04-06},
	date = {2021-03-01},
	langid = {english},
	note = {Number: 1
Publisher: Nature Publishing Group},
}

@article{jouchet_nanometric_2021,
	title = {Nanometric axial localization of single fluorescent molecules with modulated excitation},
	volume = {15},
	rights = {2021 The Author(s), under exclusive licence to Springer Nature Limited},
	issn = {1749-4893},
	url = {https://www.nature.com/articles/s41566-020-00749-9},
	doi = {10.1038/s41566-020-00749-9},
	pages = {297--304},
	number = {4},
	journaltitle = {Nature Photonics},
	author = {Jouchet, Pierre and Cabriel, Clément and Bourg, Nicolas and Bardou, Marion and Poüs, Christian and Fort, Emmanuel and Lévêque-Fort, Sandrine},
	urldate = {2021-04-06},
	date = {2021-04},
	langid = {english},
	note = {Number: 4
Publisher: Nature Publishing Group},
}

@article{cnossen_localization_2020,
	title = {Localization microscopy at doubled precision with patterned illumination},
	volume = {17},
	rights = {2019 The Author(s), under exclusive licence to Springer Nature America, Inc.},
	issn = {1548-7105},
	url = {https://www.nature.com/articles/s41592-019-0657-7},
	doi = {10.1038/s41592-019-0657-7},
	pages = {59--63},
	number = {1},
	journaltitle = {Nature Methods},
	author = {Cnossen, Jelmer and Hinsdale, Taylor and Thorsen, Rasmus Ø and Siemons, Marijn and Schueder, Florian and Jungmann, Ralf and Smith, Carlas S. and Rieger, Bernd and Stallinga, Sjoerd},
	urldate = {2021-04-06},
	date = {2020-01},
	langid = {english},
	note = {Number: 1
Publisher: Nature Publishing Group},
}

@article{gu_molecular_2019,
	title = {Molecular resolution imaging by repetitive optical selective exposure},
	volume = {16},
	rights = {2019 The Author(s), under exclusive licence to Springer Nature America, Inc.},
	issn = {1548-7105},
	url = {https://www.nature.com/articles/s41592-019-0544-2},
	doi = {10.1038/s41592-019-0544-2},
	pages = {1114--1118},
	number = {11},
	journaltitle = {Nature Methods},
	author = {Gu, Lusheng and Li, Yuanyuan and Zhang, Shuwen and Xue, Yanhong and Li, Weixing and Li, Dong and Xu, Tao and Ji, Wei},
	urldate = {2021-04-06},
	date = {2019-11},
	langid = {english},
	note = {Number: 11
Publisher: Nature Publishing Group},
}

@article{balzarotti_nanometer_2017,
	title = {Nanometer resolution imaging and tracking of fluorescent molecules with minimal photon fluxes},
	volume = {355},
	rights = {Copyright © 2017, American Association for the Advancement of Science},
	issn = {0036-8075, 1095-9203},
	url = {https://science.sciencemag.org/content/355/6325/606},
	doi = {10.1126/science.aak9913},
	pages = {606--612},
	number = {6325},
	journaltitle = {Science},
	author = {Balzarotti, Francisco and Eilers, Yvan and Gwosch, Klaus C. and Gynnå, Arvid H. and Westphal, Volker and Stefani, Fernando D. and Elf, Johan and Hell, Stefan W.},
	urldate = {2021-04-06},
	date = {2017-02-10},
	langid = {english},
	pmid = {28008086},
	note = {Publisher: American Association for the Advancement of Science
Section: Research Article},
}

@article{hirvonen_structured_2009,
	title = {Structured illumination microscopy of a living cell},
	volume = {38},
	issn = {1432-1017},
	url = {https://doi.org/10.1007/s00249-009-0501-6},
	doi = {10.1007/s00249-009-0501-6},
	pages = {807--812},
	number = {6},
	journaltitle = {European Biophysics Journal},
	shortjournal = {Eur Biophys J},
	author = {Hirvonen, Liisa M. and Wicker, Kai and Mandula, Ondrej and Heintzmann, Rainer},
	urldate = {2021-04-05},
	date = {2009-07-01},
	langid = {english},
}

@article{kubitscheck_imaging_2000,
	title = {Imaging and Tracking of Single {GFP} Molecules in Solution},
	volume = {78},
	issn = {0006-3495},
	url = {https://www.sciencedirect.com/science/article/pii/S0006349500767646},
	doi = {10.1016/S0006-3495(00)76764-6},
	pages = {2170--2179},
	number = {4},
	journaltitle = {Biophysical Journal},
	shortjournal = {Biophysical Journal},
	author = {Kubitscheck, Ulrich and Kückmann, Oliver and Kues, Thorsten and Peters, Reiner},
	urldate = {2021-04-05},
	date = {2000-04-01},
	langid = {english},
}

@article{zheng_adaptive_2017,
	title = {Adaptive optics improves multiphoton super-resolution imaging},
	volume = {14},
	rights = {2017 Nature Publishing Group, a division of Macmillan Publishers Limited. All Rights Reserved.},
	issn = {1548-7105},
	url = {https://www.nature.com/articles/nmeth.4337},
	doi = {10.1038/nmeth.4337},
	pages = {869--872},
	number = {9},
	journaltitle = {Nature Methods},
	author = {Zheng, Wei and Wu, Yicong and Winter, Peter and Fischer, Robert and Nogare, Damian Dalle and Hong, Amy and {McCormick}, Chad and Christensen, Ryan and Dempsey, William P. and Arnold, Don B. and Zimmerberg, Joshua and Chitnis, Ajay and Sellers, James and Waterman, Clare and Shroff, Hari},
	urldate = {2021-04-05},
	date = {2017-09},
	langid = {english},
	note = {Number: 9
Publisher: Nature Publishing Group},
}

@article{booth_adaptive_2014,
	title = {Adaptive optical microscopy: the ongoing quest for a perfect image},
	volume = {3},
	rights = {2014 The Author(s)},
	issn = {2047-7538},
	url = {https://www.nature.com/articles/lsa201446},
	doi = {10.1038/lsa.2014.46},
	shorttitle = {Adaptive optical microscopy},
	pages = {e165--e165},
	number = {4},
	journaltitle = {Light: Science \& Applications},
	author = {Booth, Martin J.},
	urldate = {2021-04-05},
	date = {2014-04},
	langid = {english},
	note = {Number: 4
Publisher: Nature Publishing Group},
}

@article{andresen_high-resolution_2012,
	title = {High-Resolution Intravital Microscopy},
	volume = {7},
	issn = {1932-6203},
	url = {https://journals.plos.org/plosone/article?id=10.1371/journal.pone.0050915},
	doi = {10.1371/journal.pone.0050915},
	pages = {e50915},
	number = {12},
	journaltitle = {{PLOS} {ONE}},
	shortjournal = {{PLOS} {ONE}},
	author = {Andresen, Volker and Pollok, Karolin and Rinnenthal, Jan-Leo and Oehme, Laura and Günther, Robert and Spiecker, Heinrich and Radbruch, Helena and Gerhard, Jenny and Sporbert, Anje and Cseresnyes, Zoltan and Hauser, Anja E. and Niesner, Raluca},
	urldate = {2021-04-05},
	date = {2012-12-14},
	langid = {english},
	note = {Publisher: Public Library of Science},
}

@article{zurauskas_isosense_2019,
	title = {{IsoSense}: frequency enhanced sensorless adaptive optics through structured illumination},
	volume = {6},
	rights = {\&\#169; 2019 Optical Society of America},
	issn = {2334-2536},
	url = {https://www.osapublishing.org/optica/abstract.cfm?uri=optica-6-3-370},
	doi = {10.1364/OPTICA.6.000370},
	shorttitle = {{IsoSense}},
	pages = {370--379},
	number = {3},
	journaltitle = {Optica},
	shortjournal = {Optica, {OPTICA}},
	author = {Žurauskas, Mantas and Dobbie, Ian M. and Parton, Richard M. and Phillips, Mick A. and Göhler, Antonia and Davis, Ilan and Booth, Martin J.},
	urldate = {2021-04-05},
	date = {2019-03-20},
	note = {Publisher: Optical Society of America},
}

@article{turcotte_dynamic_2019,
	title = {Dynamic super-resolution structured illumination imaging in the living brain},
	volume = {116},
	rights = {Copyright © 2019 the Author(s). Published by {PNAS}.. This open access article is distributed under Creative Commons Attribution License 4.0 ({CC} {BY}).},
	issn = {0027-8424, 1091-6490},
	url = {https://www.pnas.org/content/116/19/9586},
	doi = {10.1073/pnas.1819965116},
	pages = {9586--9591},
	number = {19},
	journaltitle = {Proceedings of the National Academy of Sciences},
	shortjournal = {{PNAS}},
	author = {Turcotte, Raphaël and Liang, Yajie and Tanimoto, Masashi and Zhang, Qinrong and Li, Ziwei and Koyama, Minoru and Betzig, Eric and Ji, Na},
	urldate = {2021-04-05},
	date = {2019-05-07},
	langid = {english},
	pmid = {31028150},
	note = {Publisher: National Academy of Sciences
Section: Biological Sciences},
}

@article{arigovindan_effect_2012,
	title = {Effect of depth dependent spherical aberrations in 3D structured illumination microscopy},
	volume = {20},
	rights = {\&\#169; 2012 {OSA}},
	issn = {1094-4087},
	url = {https://www.osapublishing.org/oe/abstract.cfm?uri=oe-20-6-6527},
	doi = {10.1364/OE.20.006527},
	pages = {6527--6541},
	number = {6},
	journaltitle = {Optics Express},
	shortjournal = {Opt. Express, {OE}},
	author = {Arigovindan, Muthuvel and Sedat, John W. and Agard, David A.},
	urldate = {2021-04-05},
	date = {2012-03-12},
	note = {Publisher: Optical Society of America},
}

@article{so_resolution_2001,
	title = {Resolution enhancement in standing-wave total internal reflection microscopy: a point-spread-function engineering approach},
	volume = {18},
	rights = {\&\#169; 2001 Optical Society of America},
	issn = {1520-8532},
	url = {https://www.osapublishing.org/josaa/abstract.cfm?uri=josaa-18-11-2833},
	doi = {10.1364/JOSAA.18.002833},
	shorttitle = {Resolution enhancement in standing-wave total internal reflection microscopy},
	pages = {2833--2845},
	number = {11},
	journaltitle = {{JOSA} A},
	shortjournal = {J. Opt. Soc. Am. A, {JOSAA}},
	author = {So, Peter T. C. and Kwon, Hyuk-Sang and Dong, Chen Y.},
	urldate = {2021-03-14},
	date = {2001-11-01},
	note = {Publisher: Optical Society of America},
}

@article{cragg_lateral_2000,
	title = {Lateral resolution enhancement with standing evanescent waves},
	volume = {25},
	rights = {\&\#169; 2000 Optical Society of America},
	issn = {1539-4794},
	url = {https://www.osapublishing.org/ol/abstract.cfm?uri=ol-25-1-46},
	doi = {10.1364/OL.25.000046},
	pages = {46--48},
	number = {1},
	journaltitle = {Optics Letters},
	shortjournal = {Opt. Lett., {OL}},
	author = {Cragg, George E. and So, Peter T. C.},
	urldate = {2021-03-14},
	date = {2000-01-01},
	note = {Publisher: Optical Society of America},
}

@article{rayleigh_investigations_1879,
	title = {Investigations in optics, with special reference to the spectroscope},
	volume = {8},
	issn = {1941-5982},
	url = {https://doi.org/10.1080/14786447908639684},
	doi = {10.1080/14786447908639684},
	pages = {261--274},
	number = {49},
	journaltitle = {The London, Edinburgh, and Dublin Philosophical Magazine and Journal of Science},
	author = {Rayleigh, Lord},
	urldate = {2020-05-05},
	date = {1879-10-01},
}

@article{sparrow_spectroscopic_1916,
	title = {On Spectroscopic Resolving Power},
	volume = {44},
	issn = {0004-637X},
	url = {http://adsabs.harvard.edu/abs/1916ApJ....44...76S},
	doi = {10.1086/142271},
	pages = {76},
	journaltitle = {The Astrophysical Journal},
	shortjournal = {The Astrophysical Journal},
	author = {Sparrow, C. M.},
	urldate = {2020-05-05},
	date = {1916-09-01},
}

@article{strohl_speed_2017,
	title = {Speed limits of structured illumination microscopy},
	volume = {42},
	doi = {10.1364/OL.42.002511},
	pages = {2511--2514},
	number = {13},
	journaltitle = {Optics Letters},
	author = {Ströhl, Florian and Kaminski, Clemens F.},
	date = {2017},
	note = {tex.date-modified= 2017-08-09 13:45:05 +0000},
}

@article{heintzmann_subdiffraction_2009,
	title = {Subdiffraction resolution in continuous samples},
	volume = {3},
	rights = {2009 Nature Publishing Group},
	issn = {1749-4893},
	url = {https://www.nature.com/articles/nphoton.2009.102},
	doi = {10.1038/nphoton.2009.102},
	pages = {362--364},
	number = {7},
	journaltitle = {Nature Photonics},
	author = {Heintzmann, Rainer and Gustafsson, Mats G. L.},
	urldate = {2019-12-08},
	date = {2009-07},
	langid = {english},
}

@article{heintzmann_laterally_1999,
	title = {Laterally modulated excitation microscopy: improvement of resolution by using a diffraction grating},
	url = {http://proceedings.spiedigitallibrary.org/proceeding.aspx?articleid=972650},
	doi = {10.1117/12.336833},
	shorttitle = {Laterally modulated excitation microscopy},
	pages = {185--196},
	journaltitle = {{BiOS} Europe'98},
	author = {Heintzmann, Rainer and Cremer, Christoph G.},
	urldate = {2016-03-20},
	date = {1999},
}

@article{heintzmann_saturated_2003,
	title = {Saturated patterned excitation microscopy with two-dimensional excitation patterns},
	volume = {34},
	issn = {0968-4328},
	url = {http://www.sciencedirect.com/science/article/pii/S0968432803000532},
	doi = {10.1016/S0968-4328(03)00053-2},
	series = {Super-Resolution},
	pages = {283--291},
	number = {6},
	journaltitle = {Micron},
	shortjournal = {Micron},
	author = {Heintzmann, Rainer},
	urldate = {2019-10-11},
	date = {2003-10-01},
}

@article{gao_noninvasive_2012,
	title = {Noninvasive Imaging beyond the Diffraction Limit of 3D Dynamics in Thickly Fluorescent Specimens},
	volume = {151},
	issn = {0092-8674},
	url = {http://www.sciencedirect.com/science/article/pii/S0092867412012226},
	doi = {10.1016/j.cell.2012.10.008},
	pages = {1370--1385},
	number = {6},
	journaltitle = {Cell},
	shortjournal = {Cell},
	author = {Gao, Liang and Shao, Lin and Higgins, Christopher D. and Poulton, John S. and Peifer, Mark and Davidson, Michael W. and Wu, Xufeng and Goldstein, Bob and Betzig, Eric},
	urldate = {2019-08-08},
	date = {2012-12-07},
}

@article{hess_ultra-high_2006,
	title = {Ultra-High Resolution Imaging by Fluorescence Photoactivation Localization Microscopy},
	volume = {91},
	issn = {0006-3495},
	url = {http://www.sciencedirect.com/science/article/pii/S0006349506721403},
	doi = {10.1529/biophysj.106.091116},
	pages = {4258--4272},
	number = {11},
	journaltitle = {Biophysical Journal},
	shortjournal = {Biophysical Journal},
	author = {Hess, Samuel T. and Girirajan, Thanu P. K. and Mason, Michael D.},
	urldate = {2015-12-14},
	date = {2006-12-01},
}

@article{manton_trispim:_2016,
	title = {{triSPIM}: light sheet microscopy with isotropic super-resolution},
	volume = {41},
	issn = {0146-9592, 1539-4794},
	url = {https://www.osapublishing.org/abstract.cfm?URI=ol-41-18-4170},
	doi = {10.1364/OL.41.004170},
	shorttitle = {{triSPIM}},
	pages = {4170},
	number = {18},
	journaltitle = {Optics Letters},
	author = {Manton, James D. and Rees, Eric J.},
	urldate = {2016-09-05},
	date = {2016-09-15},
	langid = {english},
}

@article{rust_sub-diffraction-limit_2006,
	title = {Sub-diffraction-limit imaging by stochastic optical reconstruction microscopy ({STORM})},
	volume = {3},
	rights = {© 2006 Nature Publishing Group},
	issn = {1548-7091},
	url = {http://www.nature.com/nmeth/journal/v3/n10/full/nmeth929.html},
	doi = {10.1038/nmeth929},
	pages = {793--796},
	number = {10},
	journaltitle = {Nature Methods},
	shortjournal = {Nat Meth},
	author = {Rust, Michael J. and Bates, Mark and Zhuang, Xiaowei},
	urldate = {2015-12-14},
	date = {2006-10},
	langid = {english},
}

@article{gustafsson_sevenfold_1995,
	title = {Sevenfold improvement of axial resolution in 3D wide-field microscopy using two objective lenses},
	volume = {2412},
	url = {http://dx.doi.org/10.1117/12.205334},
	doi = {10.1117/12.205334},
	pages = {147--156},
	journaltitle = {{SPIE} Proceedings},
	author = {Gustafsson, Mats G. L. and Agard, David A. and Sedat, John W.},
	urldate = {2016-05-05},
	date = {1995},
}

@article{heintzmann_saturated_2002,
	title = {Saturated patterned excitation microscopy—a concept for optical resolution improvement},
	volume = {19},
	url = {https://www.osapublishing.org/abstract.cfm?uri=josaa-19-8-1599},
	pages = {1599--1609},
	number = {8},
	journaltitle = {{JOSA} A},
	author = {Heintzmann, Rainer and Jovin, Thomas M. and Cremer, Christoph},
	urldate = {2016-03-20},
	date = {2002},
}

@article{planchon_rapid_2011,
	title = {Rapid three-dimensional isotropic imaging of living cells using Bessel beam plane illumination},
	volume = {8},
	rights = {© 2011 Nature Publishing Group, a division of Macmillan Publishers Limited. All Rights Reserved.},
	issn = {1548-7091},
	url = {http://www.nature.com/nmeth/journal/v8/n5/full/nmeth.1586.html},
	doi = {10.1038/nmeth.1586},
	pages = {417--423},
	number = {5},
	journaltitle = {Nature Methods},
	shortjournal = {Nat Meth},
	author = {Planchon, Thomas A. and Gao, Liang and Milkie, Daniel E. and Davidson, Michael W. and Galbraith, James A. and Galbraith, Catherine G. and Betzig, Eric},
	urldate = {2016-02-11},
	date = {2011-05},
}

@article{gustafsson_nonlinear_2005,
	title = {Nonlinear structured-illumination microscopy: Wide-field fluorescence imaging with theoretically unlimited resolution},
	volume = {102},
	issn = {0027-8424, 1091-6490},
	url = {http://www.pnas.org/content/102/37/13081},
	doi = {10.1073/pnas.0406877102},
	shorttitle = {Nonlinear structured-illumination microscopy},
	pages = {13081--13086},
	number = {37},
	journaltitle = {Proceedings of the National Academy of Sciences of the United States of America},
	shortjournal = {{PNAS}},
	author = {Gustafsson, Mats G. L.},
	urldate = {2016-03-21},
	date = {2005-09-13},
	pmid = {16141335},
}

@article{chen_lattice_2014,
	title = {Lattice light-sheet microscopy: Imaging molecules to embryos at high spatiotemporal resolution},
	volume = {346},
	rights = {Copyright © 2014, American Association for the Advancement of Science},
	issn = {0036-8075, 1095-9203},
	url = {http://science.sciencemag.org/content/346/6208/1257998},
	doi = {10.1126/science.1257998},
	shorttitle = {Lattice light-sheet microscopy},
	pages = {1257998},
	number = {6208},
	journaltitle = {Science},
	author = {Chen, Bi-Chang and Legant, Wesley R. and Wang, Kai and Shao, Lin and Milkie, Daniel E. and Davidson, Michael W. and Janetopoulos, Chris and Wu, Xufeng S. and Hammer, John A. and Liu, Zhe and English, Brian P. and Mimori-Kiyosue, Yuko and Romero, Daniel P. and Ritter, Alex T. and Lippincott-Schwartz, Jennifer and Fritz-Laylin, Lillian and Mullins, R. Dyche and Mitchell, Diana M. and Bembenek, Joshua N. and Reymann, Anne-Cecile and Böhme, Ralph and Grill, Stephan W. and Wang, Jennifer T. and Seydoux, Geraldine and Tulu, U. Serdar and Kiehart, Daniel P. and Betzig, Eric},
	urldate = {2016-02-08},
	date = {2014-10-24},
	pmid = {25342811},
}

@article{shao_i5s:_2008,
	title = {I5S: Wide-Field Light Microscopy with 100-nm-Scale Resolution in Three Dimensions},
	volume = {94},
	issn = {0006-3495},
	url = {http://www.ncbi.nlm.nih.gov/pmc/articles/PMC2397336/},
	doi = {10.1529/biophysj.107.120352},
	shorttitle = {I5S},
	pages = {4971--4983},
	number = {12},
	journaltitle = {Biophysical Journal},
	shortjournal = {Biophys J},
	author = {Shao, Lin and Isaac, Berith and Uzawa, Satoru and Agard, David A. and Sedat, John W. and Gustafsson, Mats G. L.},
	urldate = {2016-03-20},
	date = {2008-06-15},
	pmid = {18326649},
	pmcid = {PMC2397336},
}

@article{keller_fast_2010,
	title = {Fast, high-contrast imaging of animal development with scanned light sheet-based structured-illumination microscopy},
	volume = {7},
	rights = {© 2010 Nature Publishing Group, a division of Macmillan Publishers Limited. All Rights Reserved.},
	issn = {1548-7091},
	url = {http://www.nature.com/nmeth/journal/v7/n8/abs/nmeth.1476.html},
	doi = {10.1038/nmeth.1476},
	pages = {637--642},
	number = {8},
	journaltitle = {Nature Methods},
	shortjournal = {Nat Meth},
	author = {Keller, Philipp J. and Schmidt, Annette D. and Santella, Anthony and Khairy, Khaled and Bao, Zhirong and Wittbrodt, Joachim and Stelzer, Ernst H. K.},
	urldate = {2016-04-04},
	date = {2010-08},
	langid = {english},
}

@article{hell_breaking_1994,
	title = {Breaking the diffraction resolution limit by stimulated emission: stimulated-emission-depletion fluorescence microscopy},
	volume = {19},
	issn = {0146-9592, 1539-4794},
	url = {https://www.osapublishing.org/ol/abstract.cfm?uri=ol-19-11-780},
	doi = {10.1364/OL.19.000780},
	shorttitle = {Breaking the diffraction resolution limit by stimulated emission},
	pages = {780},
	number = {11},
	journaltitle = {Optics Letters},
	author = {Hell, Stefan W. and Wichmann, Jan},
	urldate = {2015-12-14},
	date = {1994-06-01},
	langid = {english},
}

@article{betzig_imaging_2006,
	title = {Imaging Intracellular Fluorescent Proteins at Nanometer Resolution},
	volume = {313},
	issn = {0036-8075, 1095-9203},
	url = {http://www.sciencemag.org/content/313/5793/1642},
	doi = {10.1126/science.1127344},
	pages = {1642--1645},
	number = {5793},
	journaltitle = {Science},
	shortjournal = {Science},
	author = {Betzig, Eric and Patterson, George H. and Sougrat, Rachid and Lindwasser, O. Wolf and Olenych, Scott and Bonifacino, Juan S. and Davidson, Michael W. and Lippincott-Schwartz, Jennifer and Hess, Harald F.},
	urldate = {2015-12-14},
	date = {2006-09-15},
	pmid = {16902090},
}

@article{abbe_beitrage_1873,
	title = {Beiträge zur Theorie des Mikroskops und der mikroskopischen Wahrnehmung},
	volume = {9},
	pages = {413--468},
	number = {1},
	journaltitle = {Archiv für Mikroskopische Anatomie},
	author = {Abbe, Ernst},
	date = {1873},
}

@article{gustafsson_three-dimensional_2008,
	title = {Three-Dimensional Resolution Doubling in Wide-Field Fluorescence Microscopy by Structured Illumination},
	volume = {94},
	issn = {0006-3495},
	url = {http://www.sciencedirect.com/science/article/pii/S0006349508703606},
	doi = {10.1529/biophysj.107.120345},
	pages = {4957--4970},
	number = {12},
	journaltitle = {Biophysical Journal},
	shortjournal = {Biophysical Journal},
	author = {Gustafsson, Mats G. L. and Shao, Lin and Carlton, Peter M. and Wang, C. J. Rachel and Golubovskaya, Inna N. and Cande, W. Zacheus and Agard, David A. and Sedat, John W.},
	urldate = {2017-04-26},
	date = {2008-06-15},
}

@article{muller_open-source_2016,
	title = {Open-source image reconstruction of super-resolution structured illumination microscopy data in {ImageJ}},
	volume = {7},
	rights = {2016 Nature Publishing Group},
	issn = {2041-1723},
	url = {https://www.nature.com/articles/ncomms10980},
	doi = {10.1038/ncomms10980},
	pages = {ncomms10980},
	journaltitle = {Nature Communications},
	author = {Müller, Marcel and Mönkemöller, Viola and Hennig, Simon and Hübner, Wolfgang and Huser, Thomas},
	urldate = {2017-09-12},
	date = {2016-03-21},
	langid = {english},
}

@article{abrahamsson_multifocus_2017,
	title = {Multifocus structured illumination microscopy for fast volumetric super-resolution imaging},
	volume = {8},
	rights = {© 2017 Optical Society of America},
	issn = {2156-7085},
	url = {https://www.osapublishing.org/abstract.cfm?uri=boe-8-9-4135},
	doi = {10.1364/BOE.8.004135},
	pages = {4135--4140},
	number = {9},
	journaltitle = {Biomedical Optics Express},
	shortjournal = {Biomed. Opt. Express, {BOE}},
	author = {Abrahamsson, Sara and Blom, Hans and Agostinho, Ana and Jans, Daniel C. and Jost, Aurelie and Müller, Marcel and Nilsson, Linnea and Bernhem, Kristoffer and Lambert, Talley J. and Heintzmann, Rainer and Brismar, Hjalmar},
	urldate = {2017-09-29},
	date = {2017-09-01},
}

@article{frohn_true_2000,
	title = {True optical resolution beyond the Rayleigh limit achieved by standing wave illumination},
	volume = {97},
	rights = {Copyright © The National Academy of Sciences},
	issn = {0027-8424, 1091-6490},
	url = {http://www.pnas.org/content/97/13/7232},
	doi = {10.1073/pnas.130181797},
	pages = {7232--7236},
	number = {13},
	journaltitle = {Proceedings of the National Academy of Sciences},
	shortjournal = {{PNAS}},
	author = {Frohn, Jan T. and Knapp, Helmut F. and Stemmer, Andreas},
	urldate = {2018-03-13},
	date = {2000-06-20},
	langid = {english},
	pmid = {10840057},
}

@article{shao_super-resolution_2011,
	title = {Super-resolution 3D microscopy of live whole cells using structured illumination},
	volume = {8},
	rights = {2011 Nature Publishing Group},
	issn = {1548-7105},
	url = {https://www.nature.com/articles/nmeth.1734},
	doi = {10.1038/nmeth.1734},
	pages = {1044--1046},
	number = {12},
	journaltitle = {Nature Methods},
	author = {Shao, Lin and Kner, Peter and Rego, E. Hesper and Gustafsson, Mats G. L.},
	urldate = {2018-03-16},
	date = {2011-12},
	langid = {english},
}

@article{chang_csilsfm_2017,
	title = {{csiLSFM} combines light-sheet fluorescence microscopy and coherent structured illumination for a lateral resolution below 100 nm},
	volume = {114},
	issn = {0027-8424, 1091-6490},
	url = {http://www.pnas.org/content/114/19/4869},
	doi = {10.1073/pnas.1609278114},
	pages = {4869--4874},
	number = {19},
	journaltitle = {Proceedings of the National Academy of Sciences},
	shortjournal = {{PNAS}},
	author = {Chang, Bo-Jui and Meza, Victor Didier Perez and Stelzer, Ernst H. K.},
	urldate = {2017-06-02},
	date = {2017-05-09},
	langid = {english},
	pmid = {28438995},
}

@article{wicker_non-iterative_2013,
	title = {Non-iterative determination of pattern phase in structured illumination microscopy using auto-correlations in Fourier space},
	volume = {21},
	rights = {© 2013 {OSA}},
	issn = {1094-4087},
	url = {https://www.osapublishing.org/abstract.cfm?uri=oe-21-21-24692},
	doi = {10.1364/OE.21.024692},
	pages = {24692--24701},
	number = {21},
	journaltitle = {Optics Express},
	shortjournal = {Opt. Express, {OE}},
	author = {Wicker, Kai},
	urldate = {2017-05-16},
	date = {2013-10-21},
}

@article{kner_super-resolution_2009,
	title = {Super-resolution video microscopy of live cells by structured illumination},
	volume = {6},
	rights = {© 2009 Nature Publishing Group},
	issn = {1548-7091},
	url = {http://www.nature.com/nmeth/journal/v6/n5/full/nmeth.1324.html},
	doi = {10.1038/nmeth.1324},
	pages = {339--342},
	number = {5},
	journaltitle = {Nature Methods},
	shortjournal = {Nat Meth},
	author = {Kner, Peter and Chhun, Bryant B. and Griffis, Eric R. and Winoto, Lukman and Gustafsson, Mats G. L.},
	urldate = {2017-05-07},
	date = {2009-05},
	langid = {english},
}

@article{sheppard_super-resolution_1988,
	title = {Super-resolution in Confocal Imaging},
	volume = {80},
	pages = {83--84},
	journaltitle = {Optik},
	author = {Sheppard, C. J. R.},
	date = {1988},
}

@article{song_fast_2016,
	title = {Fast structured illumination microscopy using rolling shutter cameras},
	volume = {27},
	issn = {0957-0233},
	url = {http://stacks.iop.org/0957-0233/27/i=5/a=055401},
	doi = {10.1088/0957-0233/27/5/055401},
	pages = {055401},
	number = {5},
	journaltitle = {Measurement Science and Technology},
	shortjournal = {Meas. Sci. Technol.},
	author = {Song, Liyan and Lu-Walther, Hui-Wen and Förster, Ronny and Jost, Aurélie and Kielhorn, Martin and Zhou, Jianying and {Rainer Heintzmann}},
	urldate = {2017-05-05},
	date = {2016},
	langid = {english},
}

@article{sheppard_interpretation_2016,
	title = {Interpretation of the optical transfer function: Significance for image scanning microscopy},
	volume = {24},
	rights = {© 2016 Optical Society of America},
	issn = {1094-4087},
	url = {http://www.osapublishing.org/abstract.cfm?uri=oe-24-24-27280},
	doi = {10.1364/OE.24.027280},
	shorttitle = {Interpretation of the optical transfer function},
	pages = {27280--27287},
	number = {24},
	journaltitle = {Optics Express},
	shortjournal = {Opt. Express, {OE}},
	author = {Sheppard, Colin J. R. and Roth, Stephan and Heintzmann, Rainer and Castello, Marco and Vicidomini, Giuseppe and Chen, Rui and Chen, Xudong and Diaspro, Alberto},
	urldate = {2016-11-24},
	date = {2016-11-28},
}

@article{rego_nonlinear_2012,
	title = {Nonlinear structured-illumination microscopy with a photoswitchable protein reveals cellular structures at 50-nm resolution},
	volume = {109},
	issn = {0027-8424, 1091-6490},
	url = {http://www.pnas.org/cgi/doi/10.1073/pnas.1107547108},
	doi = {10.1073/pnas.1107547108},
	pages = {E135--E143},
	number = {3},
	journaltitle = {Proceedings of the National Academy of Sciences},
	author = {Rego, E. H. and Shao, L. and Macklin, J. J. and Winoto, L. and Johansson, G. A. and Kamps-Hughes, N. and Davidson, M. W. and Gustafsson, M. G. L.},
	urldate = {2016-10-03},
	date = {2012-01-17},
	langid = {english},
}

@article{sharonov_wide-field_2006,
	title = {Wide-field subdiffraction imaging by accumulated binding of diffusing probes},
	volume = {103},
	issn = {0027-8424, 1091-6490},
	url = {http://www.pnas.org/content/103/50/18911},
	doi = {10.1073/pnas.0609643104},
	pages = {18911--18916},
	number = {50},
	journaltitle = {Proceedings of the National Academy of Sciences},
	shortjournal = {{PNAS}},
	author = {Sharonov, Alexey and Hochstrasser, Robin M.},
	urldate = {2016-06-29},
	date = {2006-12-12},
	langid = {english},
	pmid = {17142314},
}

@article{muller_image_2010,
	title = {Image Scanning Microscopy},
	volume = {104},
	url = {http://link.aps.org/doi/10.1103/PhysRevLett.104.198101},
	doi = {10.1103/PhysRevLett.104.198101},
	pages = {198101},
	number = {19},
	journaltitle = {Physical Review Letters},
	shortjournal = {Phys. Rev. Lett.},
	author = {Müller, Claus B. and Enderlein, Jörg},
	urldate = {2016-06-21},
	date = {2010-05-10},
}

@article{de_luca_re-scan_2013,
	title = {Re-scan confocal microscopy: scanning twice for better resolution},
	volume = {4},
	issn = {2156-7085, 2156-7085},
	url = {https://www.osapublishing.org/abstract.cfm?URI=boe-4-11-2644},
	doi = {10.1364/BOE.4.002644},
	shorttitle = {Re-scan confocal microscopy},
	pages = {2644},
	number = {11},
	journaltitle = {Biomedical Optics Express},
	author = {De Luca, Giulia M.R. and Breedijk, Ronald M.P. and Brandt, Rick A.J. and Zeelenberg, Christiaan H.C. and de Jong, Babette E. and Timmermans, Wendy and Azar, Leila Nahidi and Hoebe, Ron A. and Stallinga, Sjoerd and Manders, Erik M.M.},
	urldate = {2016-06-21},
	date = {2013-11-01},
	langid = {english},
}

@article{york_instant_2013,
	title = {Instant super-resolution imaging in live cells and embryos via analog image processing},
	volume = {10},
	rights = {© 2013 Nature Publishing Group, a division of Macmillan Publishers Limited. All Rights Reserved.},
	issn = {1548-7091},
	url = {http://www.nature.com/nmeth/journal/v10/n11/full/nmeth.2687.html%3Fmessage-global%3Dremove},
	doi = {10.1038/nmeth.2687},
	pages = {1122--1126},
	number = {11},
	journaltitle = {Nature Methods},
	shortjournal = {Nat Meth},
	author = {York, Andrew G. and Chandris, Panagiotis and Nogare, Damian Dalle and Head, Jeffrey and Wawrzusin, Peter and Fischer, Robert S. and Chitnis, Ajay and Shroff, Hari},
	urldate = {2016-06-21},
	date = {2013-11},
	langid = {english},
}

@article{york_resolution_2012,
	title = {Resolution doubling in live, multicellular organisms via multifocal structured illumination microscopy},
	volume = {9},
	rights = {© 2012 Nature Publishing Group, a division of Macmillan Publishers Limited. All Rights Reserved.},
	issn = {1548-7091},
	url = {http://www.nature.com/nmeth/journal/v9/n7/full/nmeth.2025.html},
	doi = {10.1038/nmeth.2025},
	pages = {749--754},
	number = {7},
	journaltitle = {Nature Methods},
	shortjournal = {Nat Meth},
	author = {York, Andrew G. and Parekh, Sapun H. and Nogare, Damian Dalle and Fischer, Robert S. and Temprine, Kelsey and Mione, Marina and Chitnis, Ajay B. and Combs, Christian A. and Shroff, Hari},
	urldate = {2016-06-21},
	date = {2012-07},
	langid = {english},
}

@article{roth_superconcentration_2016,
	title = {Superconcentration of light: circumventing the classical limit to achievable irradiance},
	volume = {41},
	issn = {0146-9592, 1539-4794},
	url = {https://www.osapublishing.org/abstract.cfm?URI=ol-41-9-2109},
	doi = {10.1364/OL.41.002109},
	shorttitle = {Superconcentration of light},
	pages = {2109},
	number = {9},
	journaltitle = {Optics Letters},
	author = {Roth, Stephan and Sheppard, Colin J. R. and Heintzmann, Rainer},
	urldate = {2016-06-16},
	date = {2016-05-01},
	langid = {english},
}

@article{roth_optical_2013,
	title = {Optical photon reassignment microscopy ({OPRA})},
	volume = {2},
	issn = {2192-2853},
	url = {http://optnano.springeropen.com/articles/10.1186/2192-2853-2-5},
	doi = {10.1186/2192-2853-2-5},
	pages = {5},
	number = {1},
	journaltitle = {Optical Nanoscopy},
	author = {Roth, Stephan and Sheppard, Colin {JR} and Wicker, Kai and Heintzmann, Rainer},
	urldate = {2016-06-16},
	date = {2013},
	langid = {english},
}

@article{betzig_breaking_1991,
	title = {Breaking the Diffraction Barrier: Optical Microscopy on a Nanometric Scale},
	volume = {251},
	rights = {1991 by the American Association for the Advancement of Science},
	issn = {0036-8075, 1095-9203},
	url = {http://science.sciencemag.org/content/251/5000/1468},
	doi = {10.1126/science.251.5000.1468},
	shorttitle = {Breaking the Diffraction Barrier},
	pages = {1468--1470},
	number = {5000},
	journaltitle = {Science},
	author = {Betzig, E. and Trautman, J. K. and Harris, T. D. and Weiner, J. S. and Kostelak, R. L.},
	urldate = {2016-03-22},
	date = {1991-03-22},
	langid = {english},
	pmid = {17779440},
}

@article{wicker_resolving_2014,
	title = {Resolving a misconception about structured illumination},
	volume = {8},
	rights = {© 2014 Nature Publishing Group, a division of Macmillan Publishers Limited. All Rights Reserved.},
	issn = {1749-4885},
	url = {http://www.nature.com/nphoton/journal/v8/n5/full/nphoton.2014.88.html},
	doi = {10.1038/nphoton.2014.88},
	pages = {342--344},
	number = {5},
	journaltitle = {Nature Photonics},
	shortjournal = {Nat Photon},
	author = {Wicker, Kai and Heintzmann, Rainer},
	urldate = {2016-01-31},
	date = {2014-05},
}

@article{li_extended-resolution_2015,
	title = {Extended-resolution structured illumination imaging of endocytic and cytoskeletal dynamics},
	volume = {349},
	rights = {Copyright © 2015, American Association for the Advancement of Science},
	issn = {0036-8075, 1095-9203},
	url = {http://science.sciencemag.org/content/349/6251/aab3500},
	doi = {10.1126/science.aab3500},
	pages = {aab3500},
	number = {6251},
	journaltitle = {Science},
	author = {Li, Dong and Shao, Lin and Chen, Bi-Chang and Zhang, Xi and Zhang, Mingshu and Moses, Brian and Milkie, Daniel E. and Beach, Jordan R. and Hammer, John A. and Pasham, Mithun and Kirchhausen, Tomas and Baird, Michelle A. and Davidson, Michael W. and Xu, Pingyong and Betzig, Eric},
	urldate = {2016-03-21},
	date = {2015-08-28},
	pmid = {26315442},
}

@article{klar_subdiffraction_1999,
	title = {Subdiffraction resolution in far-field fluorescence microscopy},
	volume = {24},
	issn = {0146-9592, 1539-4794},
	url = {https://www.osapublishing.org/abstract.cfm?URI=ol-24-14-954},
	doi = {10.1364/OL.24.000954},
	pages = {954},
	number = {14},
	journaltitle = {Optics Letters},
	author = {Klar, Thomas A. and Hell, Stefan W.},
	urldate = {2016-03-17},
	date = {1999-07-15},
	langid = {english},
}

@article{klar_fluorescence_2000,
	title = {Fluorescence microscopy with diffraction resolution barrier broken by stimulated emission},
	volume = {97},
	issn = {0027-8424, 1091-6490},
	url = {http://www.pnas.org/content/97/15/8206},
	doi = {10.1073/pnas.97.15.8206},
	pages = {8206--8210},
	number = {15},
	journaltitle = {Proceedings of the National Academy of Sciences},
	shortjournal = {{PNAS}},
	author = {Klar, Thomas A. and Jakobs, Stefan and Dyba, Marcus and Egner, Alexander and Hell, Stefan W.},
	urldate = {2016-03-17},
	date = {2000-07-18},
	langid = {english},
	pmid = {10899992},
}

@article{lal_structured_2016,
	title = {Structured illumination microscopy image reconstruction algorithm},
	volume = {{PP}},
	issn = {1077-260X},
	doi = {10.1109/JSTQE.2016.2521542},
	pages = {1--1},
	number = {99},
	journaltitle = {{IEEE} Journal of Selected Topics in Quantum Electronics},
	author = {Lal, A. and Shan, C. and Xi, P.},
	date = {2016},
}

@article{mudry_structured_2012,
	title = {Structured illumination microscopy using unknown speckle patterns},
	volume = {6},
	rights = {© 2012 Nature Publishing Group},
	issn = {1749-4885},
	url = {http://www.nature.com/nphoton/journal/v6/n5/full/nphoton.2012.83.html#ref1},
	doi = {10.1038/nphoton.2012.83},
	pages = {312--315},
	number = {5},
	journaltitle = {Nature Photonics},
	shortjournal = {Nat Photon},
	author = {Mudry, E. and Belkebir, K. and Girard, J. and Savatier, J. and Moal, E. Le and Nicoletti, C. and Allain, M. and Sentenac, A.},
	urldate = {2015-04-15},
	date = {2012-05},
	langid = {english},
}

\begin{video}
    \caption{The effect of increased illumination power on harmonic strength in saturated structured illumination microscopy.}
    \label{svid:ssim}
\end{video}

\begin{video}
    \caption{Shifting high-resolution information back to correct locations in Fourier space.}
    \label{svid:shifting}
\end{video}

\begin{video}
    \caption{Demonstration of homodyne reconstruction method as applied to scanning `SIM' as an excitation pattern.}
    \label{svid:sim_ism}
\end{video}

\end{document}